\documentclass[fleqn,usenatbib]{mnras}

\usepackage{mathptmx}

\usepackage[T1]{fontenc}
\usepackage{ae,aecompl}
\usepackage{tikz}

\usepackage{graphicx}    
\usepackage{amsmath}    
\usepackage{amssymb}    
\usepackage{gensymb}

\title[The black hole mass of NGC 4546]{Measuring the mass of the supermassive black hole of the lenticular galaxy NGC 4546}

\author[Ricci \& Steiner]{
T. V. Ricci$^{1}$\thanks{E-mail: tiago.ricci@uffs.edu.br }
J. E. Steiner$^{2}$ \\
$^{1}$Campus Cerro Largo, Universidade Federal da Fronteira Sul, RS 97900-000, Brazil\\
$^{2}$Instituto de Astronomia, Geof\'isica e Ci\^encias Atmosf\'ericas, Universidade de S\~ao Paulo, Brazil\\
}

\date{Accepted XXX. Received YYY; in original form ZZZ}

\pubyear{2020}

\begin{document}
\label{firstpage}
\pagerange{\pageref{firstpage}--\pageref{lastpage}}
\maketitle

\begin{abstract}
Most galaxies with a well-structured bulge host a supermassive black hole (SMBH) in their centre. Stellar kinematics models applied to adaptive optics (AO) assisted integral field unit observations are well-suited to measure the SMBH mass ($M_{BH}$) and also the total mass-to-light ratio [$(M/L)_{TOT}$] and possible anisotropies in the stellar velocity distribution in the central region of galaxies. In this work, we used new AO assisted Near-Infrared Integral Field Spectrometer (NIFS) observations and also photometric data from the Hubble Space Telescope Legacy Archive of the galaxy NGC 4546 in order to determine its SMBH mass. To do this, we applied the Jeans Anisotropic Modelling (JAM) method to fit the average second velocity moment in the line of sight $(\overline{v^2_{los}})$ of the stellar structure. In addition, we also obtained $(M/L)_{TOT}$ and the classical anisotropy parameter $\beta_z$=1--($\sigma_z$/$\sigma_R$)$^2$ for this object within a field of view of 200$\times$200 pc$^2$. Maps of the stellar radial velocity and of the velocity dispersion were built for this galaxy using the penalized pixel fitting ({\sc ppxf}) technique. We applied the Multi Gaussian Expansion  procedure to fit the stellar brightness distribution.  Using JAM, the best-fitting model for $\overline{v^2_{los}}$ of the stellar structure was obtained with $(M/L)_{TOT}$ = 4.34$\pm$0.07 (Johnson's R band), $M_{BH}$ = (2.56$\pm$0.16)$\times$10$^8$ M$_\odot$ and $\beta_z$ = --0.015$\pm$0.03 (3$\sigma$ confidence level).  With these results, we found that NGC 4546 follows the $M_{BH}$ $\times$ $\sigma$ relation. We also measured the central velocity dispersion within a radius of 1 arcsec of this object as $\sigma_c$ = 241$\pm$2 km s$^{-1}$.     

\end{abstract}




\section{Introduction} \label{sec:introduction}

It is well established that the central region of most galaxies with bulges hosts a supermassive black hole (SMBH). Observational evidence for the presence of an SMBH include the existence of active galactic nuclei (AGNs; \citealt{2008ARA&A..46..475H, 2013peag.book.....N}), gravitational signatures on stellar and gas kinematics  \citep{1998AJ....115.2285M, 2005SSRv..116..523F, 2013ARA&A..51..511K} and, more recently, the first direct imaging of the event horizon of M87$^*$ \citep{2019ApJ...875L...1E}. Moreover, the relations between the mass of SMBHs ($M_{BH}$) and global parameters of galaxies, such as the velocity dispersion \citep{2000ApJ...539L...9F, 2000ApJ...539L..13G, 2009ApJ...698..198G, 2016ApJ...818...47S} and the stellar mass of the bulge \citep{1998AJ....115.2285M, 2004ApJ...604L..89H} suggest a co-evolution between the SMBH and its host (see e.g. \citealt{2013ARA&A..51..511K}  for a review).

Stellar dynamical models are an important tool to determine SMBH masses. However, to obtain a reliable measurement for $M_{BH}$, it is important to use both spectroscopic and photometric data whose spatial resolution is high enough to resolve the gravitational sphere of influence of the SMBH \citep{2003ApJ...583...92G}. In addition, these authors also suggested that the use of a modelling machinery that is more general, such as the Schwarzschild's orbit superposition method \citep{1979ApJ...232..236S} is important to avoid an unbiased measurement for $M_{BH}$. \citet{2009MNRAS.399.1839K} showed that two sets of integral field unit (IFU) data are necessary when using the Schwarzschild's technique: high resolution (small FOV) to probe the sphere-of-influence radius ($R_{soi}$), and large field, natural seeing data to constrain the large-scale orbits. \citet{2014ApJ...782...39T} also pointed out the importance of using IFU data to constrain the distribution of stellar orbits in the central region of galaxies. On the other hand, \citet{2008MNRAS.390...71C} developed a way to calculate the stellar velocity moments by using the Jeans axisymmetric formalism \citep{1922MNRAS..82..122J, 1987gady.book.....B} which is completely determined by the galaxy inclination, the mass distribution and the classic anisotropy parameter $\beta_z$ = 1 -- ($\sigma_z$/$\sigma_R$)$^2$, where $\sigma_z$ and $\sigma_R$ are the velocity dispersions in the $z$ and $R$-directions, respectively. The Jeans Anisotropic Modelling (JAM) technique is based on the observational fact that most of the fast-rotating early-type galaxies are axisymmetric and that their velocity dispersion ellipses are aligned with the cylindrical coordinates ($R$, $\phi$, $z$) and flattened along the symmetry axis $z$ \citep{2007MNRAS.379..418C, 2016ARA&A..54..597C}, i.e. $\sigma_z \leq \sigma_R$. The JAM procedure has been used to measure $M_{BH}$, $\beta_z$ and the total mass-to-light ratio $(M/L)_{TOT}$ in the central region of galaxies using adaptive optics (AO) assisted IFU observations, allowing an unbiased measurement of $M_{BH}$ when compared to more general models \citep{2015MNRAS.450..128D, 2018MNRAS.477.3030K, 2019A&A...625A..62T}.  

\begin{figure*}
\begin{tikzpicture}
    \begin{scope}[xshift=-5.5cm, yshift=0.4cm]
    \node {\includegraphics[scale=0.45]{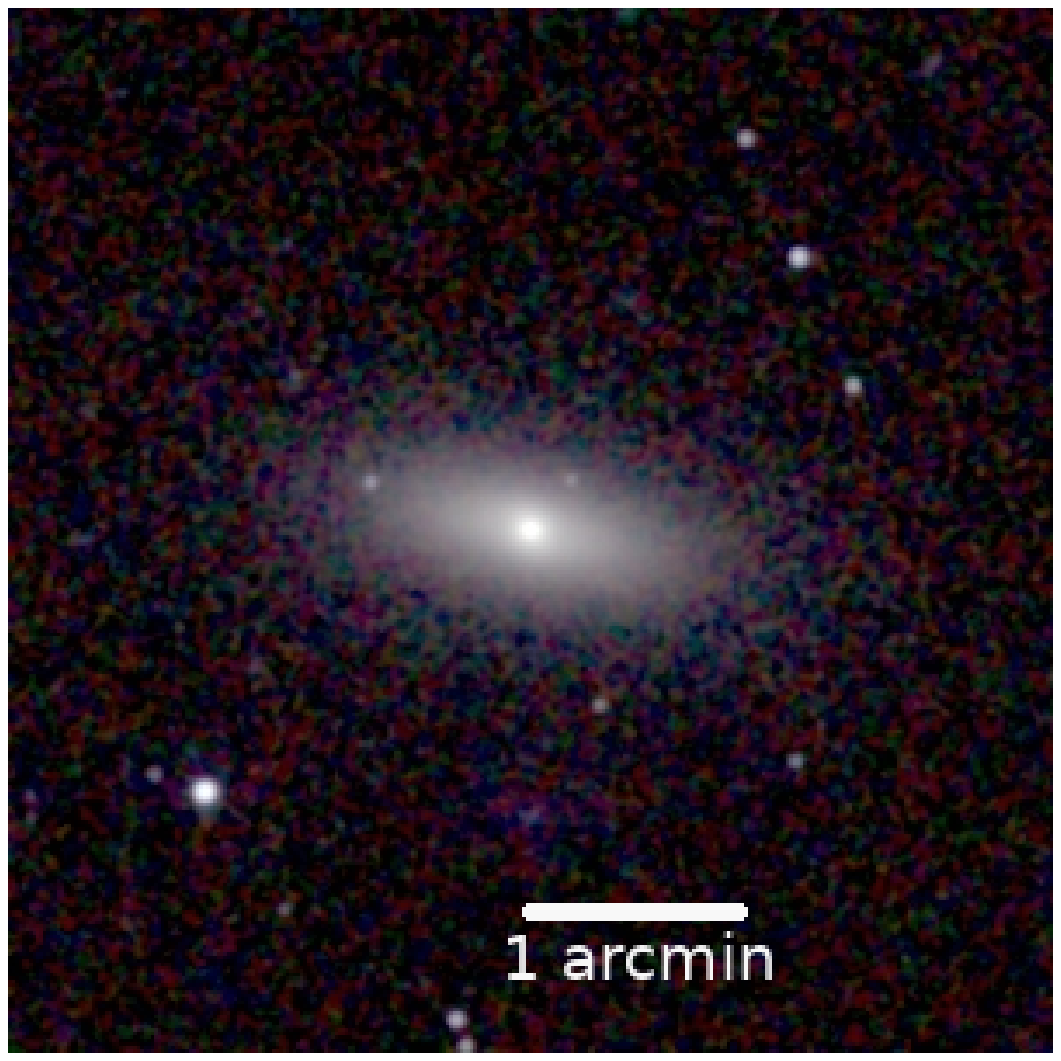}};
    \end{scope}
    \begin{scope}[xshift=0.0cm]
    \node {\includegraphics[scale=0.5]{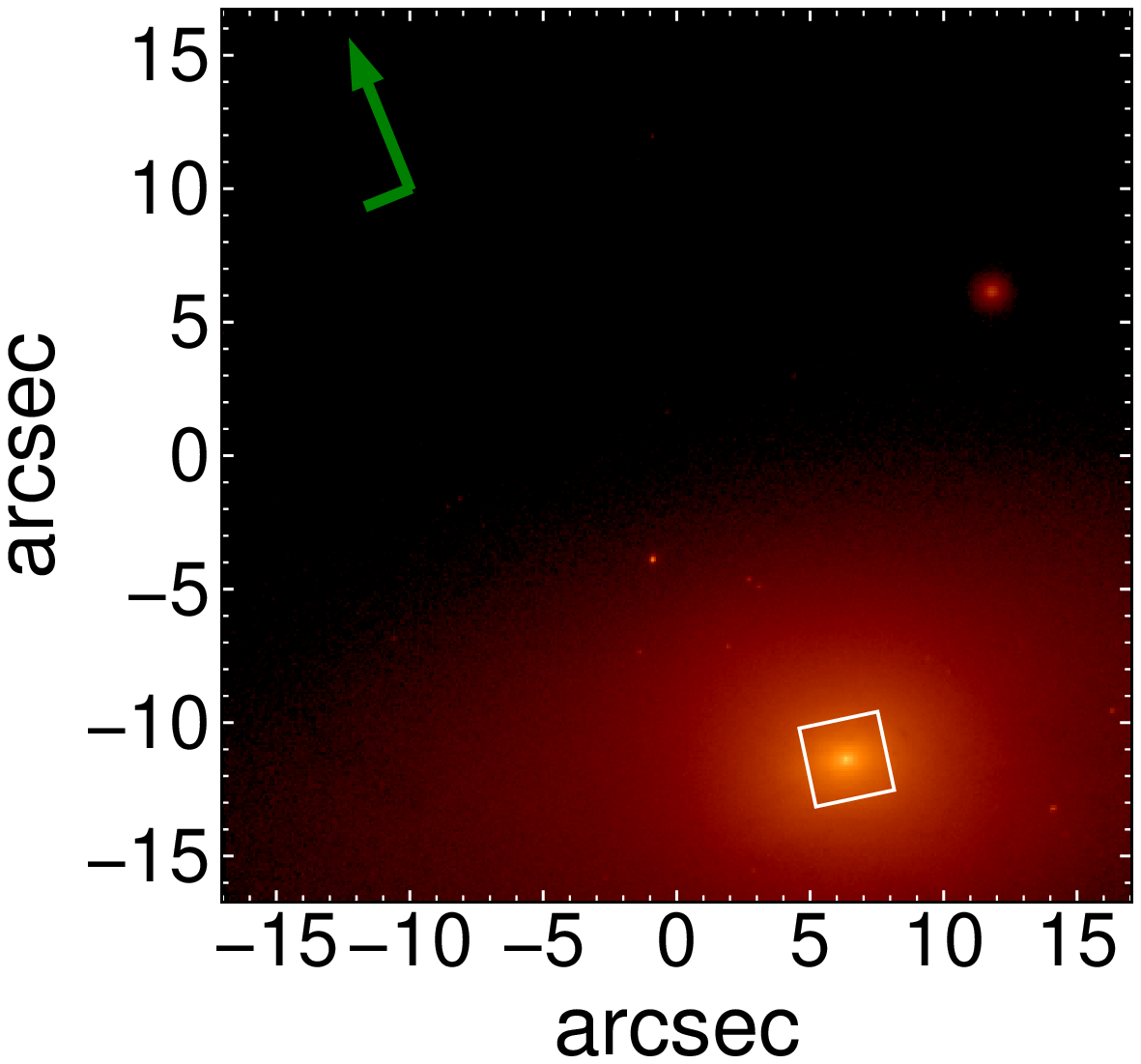}}; 
    \end{scope}
    \begin{scope}[xshift=6.0cm]
    \node {\includegraphics[scale=0.5]{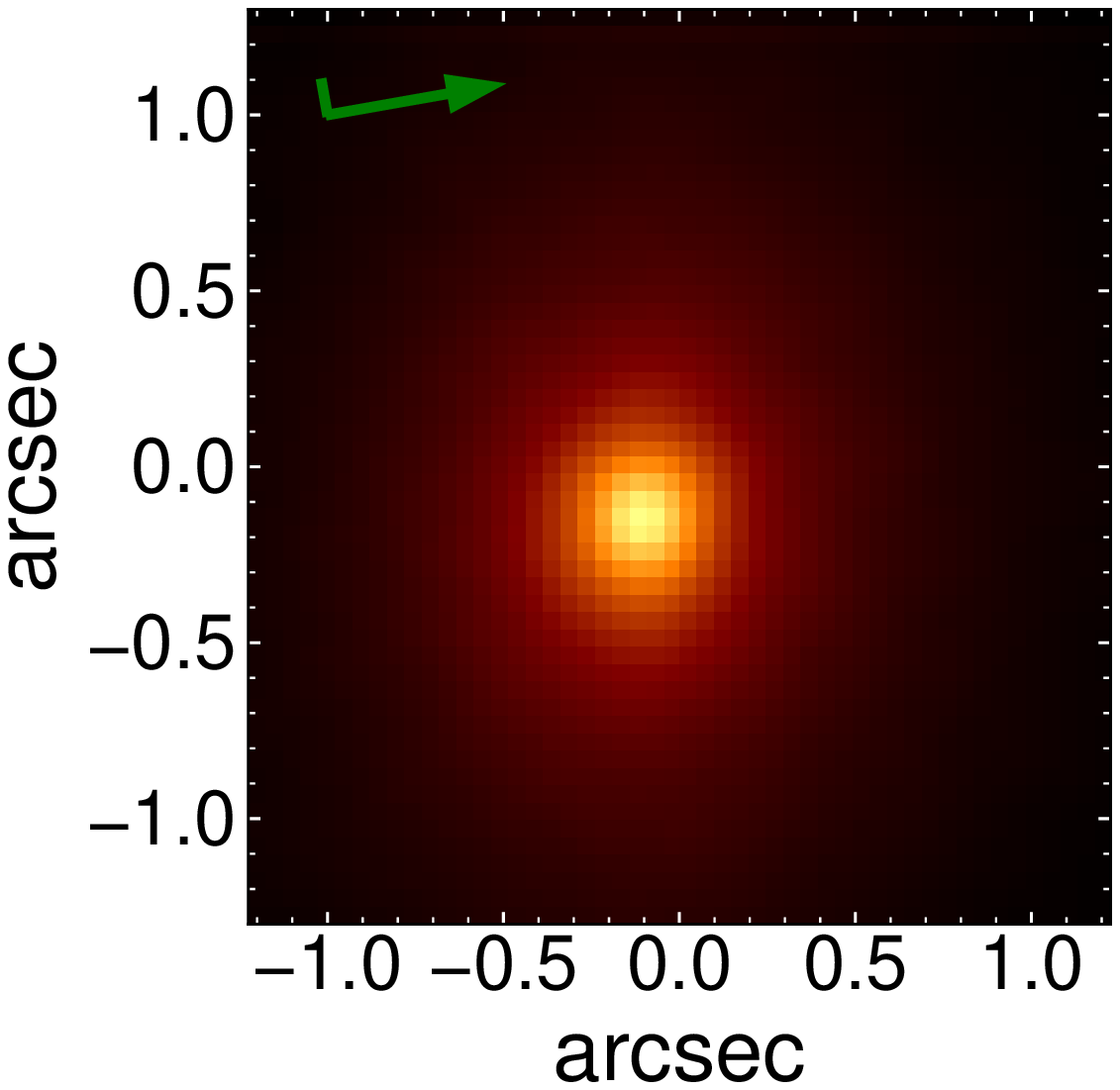}}; 
    \end{scope}

    \end{tikzpicture}
\caption{Left: composed JHK image of NGC 4546 from 2MASS \citep{2006AJ....131.1163S}. In this image, north is up and east is to the left. Centre: HST/WFPC2 image of NGC 4546, taken with the F606W filter. The solid white square represents the FOV of the NIFS observation used in this work. Right: average image of the NIFS data cube of NGC 4546, extracted between 2.24 and 2.26 $\mu$m. For this galaxy, 1 arcsec = 68 pc. The green arrow is the north direction, with east to the left. In the 2MASS image, north is up and east is to the left. \label{fig:ngc4546} }
\end{figure*}

The main goal of this paper is to use the JAM method to measure the parameters $M_{BH}$, $(M/L)_{TOT}$ and $\beta_z$ of the central region of the galaxy NGC 4546. To do this, we obtained AO assisted IFU observations from the Near-Infrared Integral Field Spectrometer (NIFS), installed on the Gemini North Telescope, and also archive photometric observations performed with the Wide Field and Planetary Camera 2 (WFPC2), installed on the Hubble Space Telescope (HST). NGC 4546 (see Fig. \ref{fig:ngc4546}) is an SB0 galaxy \citep{1991trcb.book.....D} with magnitude $M_k$ = --23.30 (\citealt{2006AJ....131.1163S}, K band) and is located at a distance of 14 Mpc (\citealt{2013AJ....146...86T}; 1 arcsec = 68 pc). It hosts a type 1 LINER AGN at the centre \citep{2014MNRAS.440.2442R}. This object was also classified as a fast rotator by \citet{2011MNRAS.414..888E}. Gas discs in molecular, neutral, and ionized forms are all counter rotating with respect to the stellar disc \citep{1987ApJ...318..531G, 1991MNRAS.248..544B, 1994AJ....108.1633S, 2006MNRAS.366.1151S, 2014MNRAS.440.2419R}. \citet{2013MNRAS.432.1709C} used JAM to model the stellar kinematics of NGC 4546 using a SAURON data cube of this galaxy from the ATLAS$^{3D}$ project \citep{2011MNRAS.413..813C}. Although their procedure was very important to determine global parameters for the stellar dynamics of NGC 4546, such as $(M/L)_{TOT}$ and $\beta_z$ within 1 effective radius (1 $R_{eff}$ = 22 arcsec, \citealt{2013MNRAS.432.1709C}), the spatial resolution of the data cubes from the ATLAS$^{3D}$ project is not high enough to resolve the sphere-of-influence radius of this object. Assuming $R_{soi} \, = \, GM_{BH} / \sigma_e^2$, where $\sigma_e$ is the velocity dispersion within 1 effective radius and using the $M_{BH}$ $\times$ $\sigma$ relation of \citet{2016ApJ...818...47S}, we calculate $R_{soi}$ = 18 pc (0.26 arcsec) for  $\sigma_e$ = 188 km s$^{-1}$ \citep{2013MNRAS.432.1709C}. We will show that $R_{soi}$ is resolved by our NIFS observations. Thus, this paper will complement the JAM results of \citet{2013MNRAS.432.1709C} for NGC 4546, but for the vicinity of its SMBH, i.e. using AO assisted IFU data within a field of view (FOV) of 200$\times$200 pc$^2$.

\section{Observations and data reduction} \label{sec:observations_data_reduction} 

Below, we present the HST/WFPC2 and the Gemini/NIFS observations of NGC 4546. A summary of both data sets is presented in Table \ref{tab:journal_observations}.  

\begin{table*}
    \centering
    \caption{Journal of observations. For the HST/WFPC2 instrument, the wavelength range is the effective range. The total bandwidth for the F606W filter is 1500 \AA\ for a central wavelength of 5860 \AA. The PSF of the NIFS observations is well described by a sum of two concentric Gaussian functions. Thus, we present the FWHM of the narrow (N) and the broad (B) components. The intensity of the narrow component Int$_N$ = 0.51. For the broad component, the intensity Int$_B$ = 1 -- Int$_N$ = 0.49, since the PSF must be normalized. \label{tab:journal_observations}}
    \begin{tabular}{cccccc}
    \hline
    Telescope/instrument & Observation date & Filter/grating & Spectral resolution ($\sigma$) & Wavelength range & Spatial resolution (FWHM)  \\
    &&&&& (arcsec) \\
    \hline
    HST/WFPC2 & 1994 May 16 & F606W & $-$ & 5100 -- 6600 \AA\ & 0.08 \\
    Gemini/NIFS & 2019 May 20 & K & 20 km s$^{-1}$ & 2.0--2.4 $\mu$m & 0.21 (N), 0.65 (B)  \\ 
    \hline
    \end{tabular}

\end{table*}

\subsection{HST/WFPC2} \label{sec:hst_observations}

NGC 4546 was observed with HST/WFPC2 on 1994 May 16 (proposal ID 5446). Only the F606W filter was used in the observations. We obtained level 2 data from the Hubble Legacy Archive website\footnote{https://hla.stsci.edu/}. It means that the two exposures that were taken from this object were bias subtracted, dark corrected, flat-fielded and reconstructed using the Drizzle method \citep{2002PASP..114..144F}. The sky was subtracted using a constant level to set the background of the combination of both exposures close to zero. We used only the Planetary Camera (PC) chip in this work, which has a spatial scale of 0.05 arcsec within a FOV of 35 $\times$ 35 arcsec$^2$. We calculated the point spread function (PSF) of the PC exposure with the {\sc tiny tim} software \citep{2011hst..psf} and we estimate the spatial resolution of these observations as 0.08 arcsec (full width at half-maximum, FWHM). We show the final image used in this work in Fig. \ref{fig:ngc4546}.

\subsection{Gemini/NIFS} \label{sec:nifs_observations}

NGC 4546 was observed on 2019 May 20 (programme GN-2019A-Q-207) with NIFS \citep{2003SPIE.4841.1581M} together with the AO system ALTAIR. Such a combination produces 3D Spectroscopy near-infrared data with an FOV of 3.0$\times$3.0 arcsec$^2$ and spatial resolutions of $\sim$ 0.1 arcsec. The K grating was used in the observations, which covered a spectral range of 2.0--2.4 $\mu$m. Eight exposures on the object (O) and four exposures on the blank sky (S) were taken, with an integration time of 500 s each, in the following sequence: OOSOOSOOSOOS. The natural seeing for the observations (i.e. without AO correction) was estimated to be $\sim$ 0.45 arcsec. 

We used the standard Gemini packages available in {\sc pyraf} to reduce the data. First, we subtracted the sky emission from the science exposures. Then, we applied flat-field and bad pixel corrections in the data. After this, we performed a wavelength calibration using an ArXe lamp exposure. Spatial distortions were corrected using a Ronchi calibration mask. To remove telluric lines, two A0V stars were observed before (HIP 54849) and after (HIP 61318) the science exposures. The observations of these standard stars were reduced using the same procedures mentioned above. The telluric templates were built by first normalizing the continuum of the spectra, followed by the removal of the intrinsic absorption lines of the objects. We also fitted a blackbody curve to the integrated spectra of these stars in order to perform a flux calibration to the data. The next step was to create data cubes for each science exposure, all of them with a spatial scale of 0.05 arcsec. Since the observations were made with spatial dithering, we used an Interactive Data Language (IDL) algorithm developed by us to set the centre of the galaxy to the same position in the FOV in all data cubes. Finally, the final data cube was the result of a median combination of these science exposures. The spectral resolution $\sigma_i$ = 20 km s$^{-1}$, as estimated with the ArXe lamp exposure. Following \citet{2009MNRAS.399.1839K}, we described the PSF of the data cube by a sum of two concentric Gaussian functions. We found FWHM$_N$ = 0.21 arcsec and FWHM$_B$ = 0.65 arcsec for the narrow and broad components, respectively (see Section \ref{sec:jam} for more details on the determination of the PSF). Using only the narrow component of the PSF, we estimated the Strehl ratio $R_{str}$ $\sim$ 0.15, assuming that the diffraction-limited Gaussian PSF for the NIFS instrument has  FWHM $\sim$ 0.07 arcsec for $\lambda$ = 2.25 $\mu$m. 

\begin{figure}
    \centering
    \includegraphics[scale=0.5]{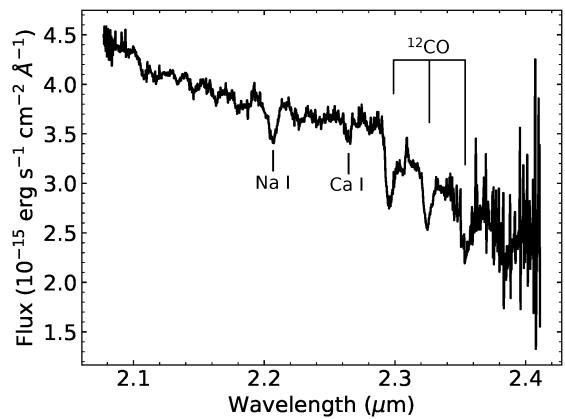}
    \caption{Spectrum extracted from the central region of the NIFS data cube of NGC 4546.}
    \label{fig:central_spectrum}
\end{figure}

Following \citet{2014MNRAS.438.2597M}, additional treatment techniques were also applied to the science data cubes. First, we removed high-frequency spatial noise using a Butterworth filter with a cut-off frequency of 0.27 F$_{NY}$, where F$_{NY}$ is the Nyquist frequency, and a filter order n = 2. In the end, we subtracted the low-frequency instrumental fingerprints using Principal Component Analysis (PCA) Tomography \citep{2009MNRAS.395...64S}. An average image of the final data cube, extracted between 2.24 and 2.26 $\mu$m, is shown in Fig. \ref{fig:ngc4546}. A representative spectrum of the central region of NGC 4546 is shown in Fig. \ref{fig:central_spectrum}. 

\section{Results} \label{sec:results}

\subsection{Stellar brightness distribution} \label{sec:stellar_brightness_distribution}

To determine the stellar surface brightness of NGC 4546, we applied the Multi Gaussian Expansion procedure (MGE, \citealt{1994A&A...285..723E}), as implemented by \citet{2002MNRAS.333..400C}, in the HST/WFPC2 image of NGC 4546. This technique fits the stellar brightness of a galaxy as a sum of 2D Gaussian functions. With the results of this parametrization, the spatial deconvolution by a PSF and the deprojection from a stellar surface brightness to a luminosity density distribution of the galaxy may be performed using simple analytic functions. In Section \ref{sec:jam}, we will use these results to calculate the stellar gravitational potential in the central region of NGC 4546.

As mentioned in Section \ref{sec:hst_observations}, HST/WFPC2 data are available for NGC 4546 for one photometric band only. Thus we did not perform any extinction correction caused by dust that may be present in NGC 4546. Although there are pieces of evidence of a dust lane along the projected major-axis of NGC 4546 \citep{1991MNRAS.248..544B, 2011ApJS..197...21H, 10.1093/mnras/staa392}, the influence of this structure in the total light (V band) of the galaxy is $\sim$ 1 \% \citep {1991MNRAS.248..544B}. Thus, any error on the stellar mass density distribution caused by the internal extinction of NGC 4546 must be small. For the foreground extinction, we adopted A$_{F606W}$ = 0.082 \citep{2011ApJ...737..103S}. 

\begin{figure*}
    \hspace{-1.5cm}
    \includegraphics[scale=0.55]{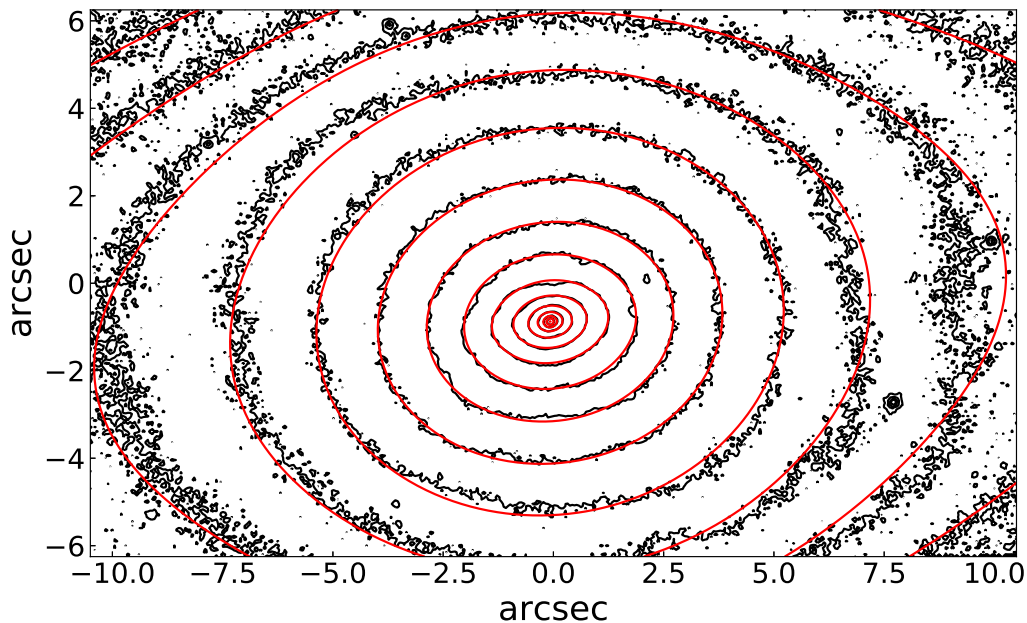}
    \caption{Comparison between the HST data, shown in black, and the MGE models, shown in red. The flux levels are normalized by the central surface brightness of the HST image and the contours are separated by a step of 0.5 mag arcsec$^{-2}$. }
    \label{fig:mge_results}
\end{figure*}

We first used the MGE technique to fit the PSF of the HST image, assuming that the 2D Gaussian functions are circular. The result is a set of four Gaussian functions whose properties (total counts $G_j$ and dispersion $\sigma_j$ for each component $j$) are shown in Table \ref{tab:mge_psf}. Note that the sum of the intensities of all Gaussian functions is equal to one since the PSF must be normalized. After this, we applied MGE to the image of NGC 4546 assuming an axisymmetric distribution, i.e. all position angles (PAs) related to the major axis of the fitted 2D Gaussian functions are the same (PA$_{maj}$ = 75$^\circ$). The parameters that describe each Gaussian function $j$ are the central surface brightness ($I_{Rj}$), the dispersion $\sigma_j$ along the major axis and the observed axial ratio $q_j$ ($q_j$ $\leq$ 1, with $q_j$ = 1 associated with a circular 2D Gaussian function). To convert from image counts to Cousins's R magnitude, we first found the total ST magnitude related to each Gaussian function following the procedure suggested in the HST/WFPC2 Data Handbook \citep{2002hstwfpc2hand.book.....O}. Then, we used a colour factor of 0.69 for a K0V star and a solar absolute magnitude in the R band $M_{R\odot}$ = 4.43 (Vega mag, \citealt{2018ApJS..236...47W}). The fitted parameters $I_{Rj}$, $\sigma_j$, and $q_j$ for each 2D Gaussian function found by the MGE procedure are shown in Table \ref{tab:mge_ngc4546}. Finally, a comparison between the HST observations and the MGE results is shown in Fig. \ref{fig:mge_results}. 

\begin{table}
    \centering
    \caption{MGE results for the PSF of the HST/WFPC2 image of NGC 4546. Each Gaussian function ($j$) used in the expansion is circular with a total count $G_j$, with $\sum_{j=1}^4 G_j$ = 1. The parameter $\sigma_j$ corresponds to the dispersion of the 2D Gaussian functions. \label{tab:mge_psf}}
    \begin{tabular}{ccc}
         \hline 
         $j$ & $G_j$ (counts) & $\sigma_j$ (arcsec) \\
         \hline 
         1 & 0.23 & 0.019 \\
         2 & 0.55 & 0.054 \\
         3 & 0.11 & 0.123 \\
         4 & 0.11 & 0.360 \\
         \hline 
    \end{tabular}
    
\end{table}

\begin{table}
    \centering
    \caption{MGE results for the HST/WFPC2 image of NGC 4546. The parameters presented below are the central surface brightness in the R band ($I_{Rj}$), the dispersion along the major axis of the Gaussian functions ($\sigma_j$) and the observed axial ratio ($q_j$). \label{tab:mge_ngc4546}}
    \begin{tabular}{cccc}
    \hline
    $j$ & $I_{Rj}$ (10$^3$ L$_{\odot}$ pc$^{-2}$) & $\sigma_j$ (arcsec) & $q_j$ \\
    \hline
    1&390 & 0.034 & 1 \\
    2&118 & 0.070 & 0.52 \\
    3&57.3 & 0.119 & 0.82 \\
    4&44.6 & 0.222 & 0.68 \\
    5&22.9 & 0.373 & 0.78 \\
    6&17.4 & 0.686 & 0.57 \\
    7&11.9 & 1.13 & 0.78 \\
    8&6.85 & 2.07 & 0.79 \\
    9&3.17 & 3.35 & 0.96 \\
    10&0.772 & 5.31 & 0.92 \\
    11&1.63 & 11.7 & 0.44 \\

    \hline
    \end{tabular}
\end{table}

In order to determine the average central logarithmic slope of the stellar density distribution of this object, we assumed a spherical distribution to deproject the MGE results shown in Table \ref{tab:mge_ngc4546}. We found that the density profile of the central region of this galaxy is well described by a power law ($\rho(r) \propto r^{-\gamma}$), with $\gamma$ $\sim$ 1.8. Such a result is in agreement with other measurements performed for other elliptical and lenticular galaxies in the local Universe (see e.g. \citealt{2013MNRAS.432.1709C}). 

\subsection{Stellar kinematics} \label{sec:stellar_kinematics}

\begin{figure}
    \includegraphics[scale=0.25]{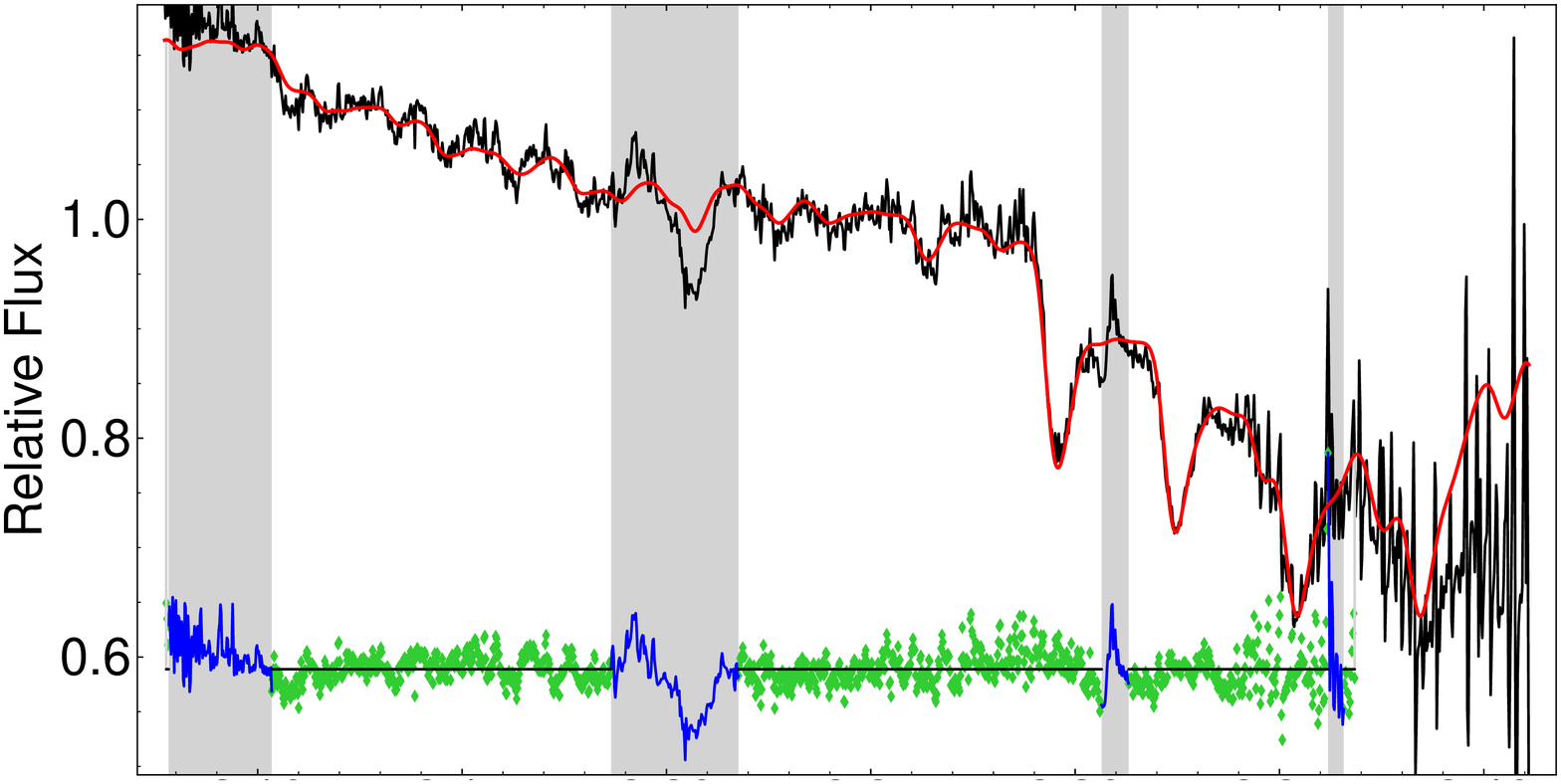}
    \includegraphics[scale=0.25]{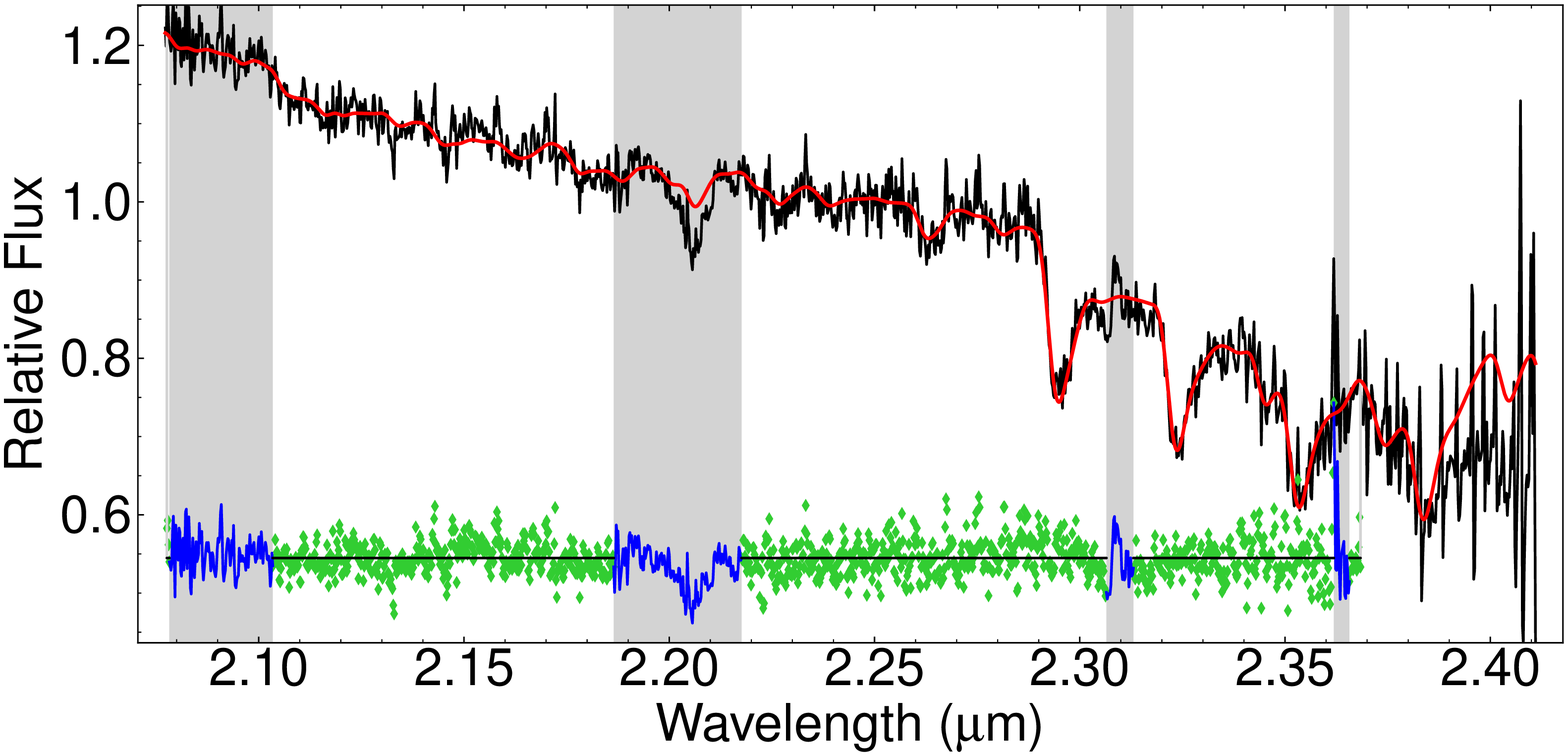}
    \caption{Fitting results of the stellar component from the data cube of NGC 4546 using {\sc ppxf}. Top: spectrum of the central spaxel. Bottom: spectrum related to a position located in the lower left region of the FOV. The observed spectra are shown in black, the fitted stellar template is shown in red and the residuals are shown in green. The grey areas correspond to the spectral regions that were masked before the fitting procedure, with the blue lines corresponding to the residuals in these ranges. \label{fig:ppxf_examples}}
\end{figure}

To extract the stellar kinematic information from the central region of NGC 4546, we used the penalized pixel fitting ({\sc ppxf}) procedure \citep{2004PASP..116..138C, 2017MNRAS.466..798C}. This software fits the stellar component of a galaxy as a combination of simple stellar population spectra convolved with a Gauss--Hermite (GH) function. The first four moments of the GH function are the radial velocity $V_{RADIAL}$, the velocity dispersion $\sigma_*$ and the GH moments $h_3$ and $h_4$, which are related to asymmetric (skewness) and symmetric (kurtosis) deviations from a Gaussian function, respectively. However, since only the $V_{RADIAL}$ and $\sigma_*$ maps are needed to compare the observations with the stellar velocity moments that are calculated with JAM (see Section \ref{sec:jam}), we fitted the line-of-sight velocity distribution (LOSVD) in the whole FOV of NGC 4546 using only a Gaussian function. This procedure was performed in the spectral range between 2.10 and 2.36 $\mu$m using the stellar library of \citet{2009ApJS..185..186W}, composed by late-type stars (G, K, M) that were observed with NIFS. We also used an additive polynomial of order 4 in the fitting procedure to consider low-frequency differences between the observed spectra and the stellar template. In Fig. \ref{fig:ppxf_examples}, we show the fitting results for the central spaxel and a location in the lower left region of the FOV. Note that we masked the Na I$\lambda$2.21 $\mu$m line since this feature is underestimated by the final template. One possible reason for this mismatch in early-type galaxies, as is the case of NGC 4546, is a combination between [Na/Fe] enhancement effects and a bottom-heavy initial mass function \citep{2017MNRAS.472..361R}. We also masked the spectral regions where the telluric absorption correction and the sky subtraction were not effective ($\lambda <$ 2.10 $\mu$m and $\lambda >$ 2.36 $\mu$m). We used the residuals to estimate the signal-to-noise (S/N) across the FOV. We found that within the spectral range used in the fitting procedure, 95 \% of all spectra of the data cube have S/N $\geq$ 40, and in 99.5 \% of all spectra, S/N $\geq$ 32.

\begin{figure}
    \includegraphics[scale=0.32]{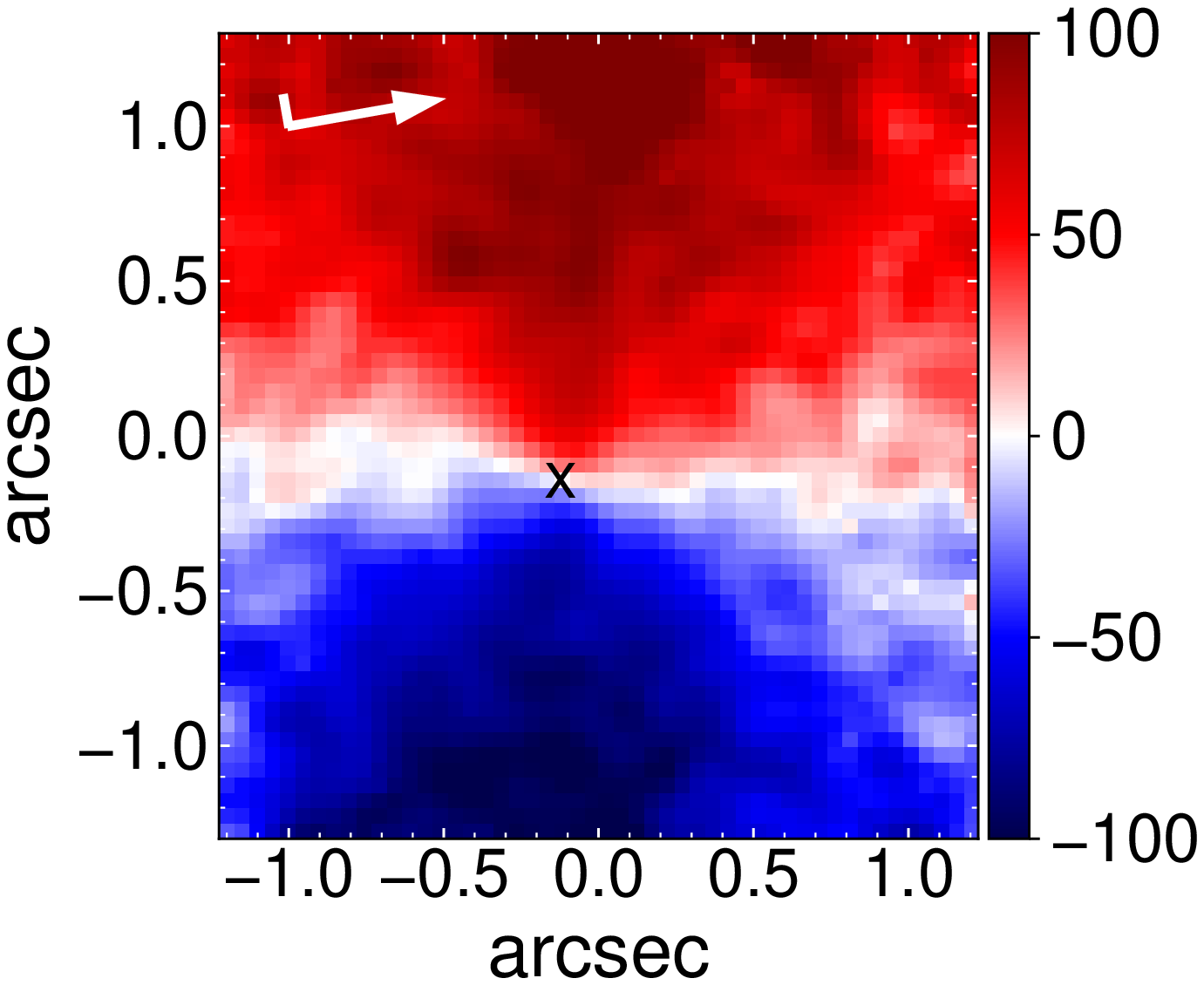}
    \includegraphics[scale=0.32]{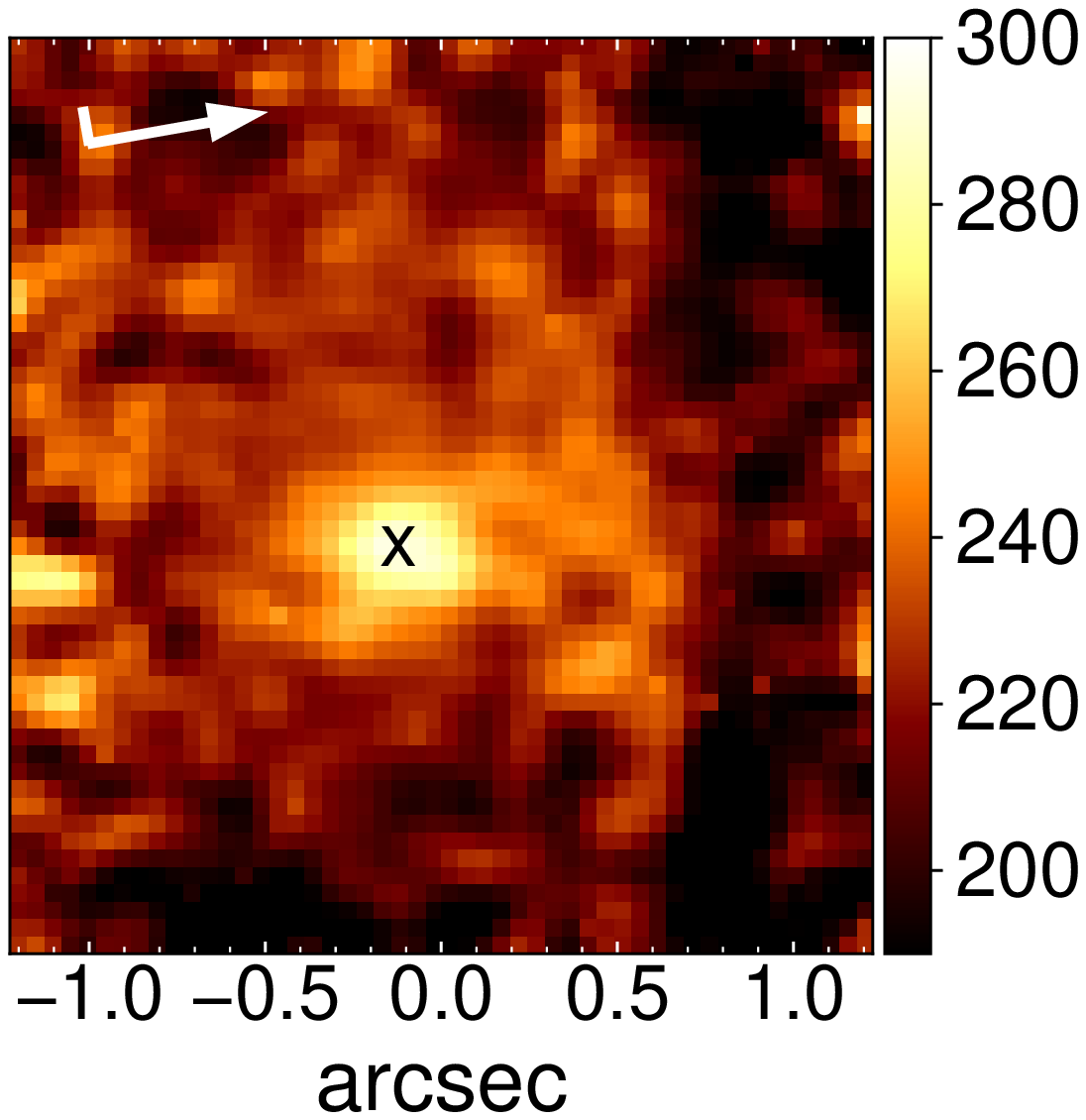}
    \caption{Stellar kinematic moments of the central region of NGC 4546. In these maps, the LOSVD was extracted using only a Gaussian function. Left: radial velocity, in km s$^{-1}$, with respect to the heliocentric velocity of NGC 4546 (V$_{HEL}$ = 1057 km s$^{-1}$). Right: velocity dispersion (in km s$^{-1}$). \label{fig:first_kinematic_results}} 
\end{figure}

We present the maps of $V_{RADIAL}$ and $\sigma_*$ of NGC 4546 in Fig. \ref{fig:first_kinematic_results}. The radial velocities in this map are shown with respect to the heliocentric velocity of the object (1057 km s$^{-1}$, \citealt{2011MNRAS.413..813C}). One may see that the stellar component of this galaxy has a clear rotation around its nuclear region. The line of nodes of the rotating structure has a PA = 76$\pm$2$^\circ$. This result is in agreement with what was measured by \citet{2016MNRAS.463.3860R} using a Gemini Multi-Object Spectrograph (GMOS) IFU data cube. The velocity dispersion $\sigma_*$ in each spaxel was calculated as $\sigma_*^2 = \sigma_m^2 - \sigma_i^2 + \sigma_b^2$, where $\sigma_m$ is the dispersion measured with {\sc ppxf}, $\sigma_i$ is the instrumental resolution and $\sigma_b$ is the template resolution. According to \citet{2009ApJS..185..186W}, $\sigma_b$ = 18 km s$^{-1}$. 

We estimated the errors of the kinematic moments by employing a Monte Carlo simulation. For each spaxel of the data cube, we added random Gaussian noise to the best-fitting stellar template spectrum and then we fitted the profile of the absorption lines, setting the bias parameter to zero. We repeated this procedure 100 times. The errors associated with $V_{RADIAL}$ and $\sigma_*$ were then calculated as the standard deviation of these measurements in the simulation. The average errors of the parameters across the data cube are $\Delta_{VRADIAL}$ = 7 km s$^{-1}$ and $\Delta_\sigma*$ = 9 km s$^{-1}$.

\begin{figure}
    \includegraphics[scale=0.31]{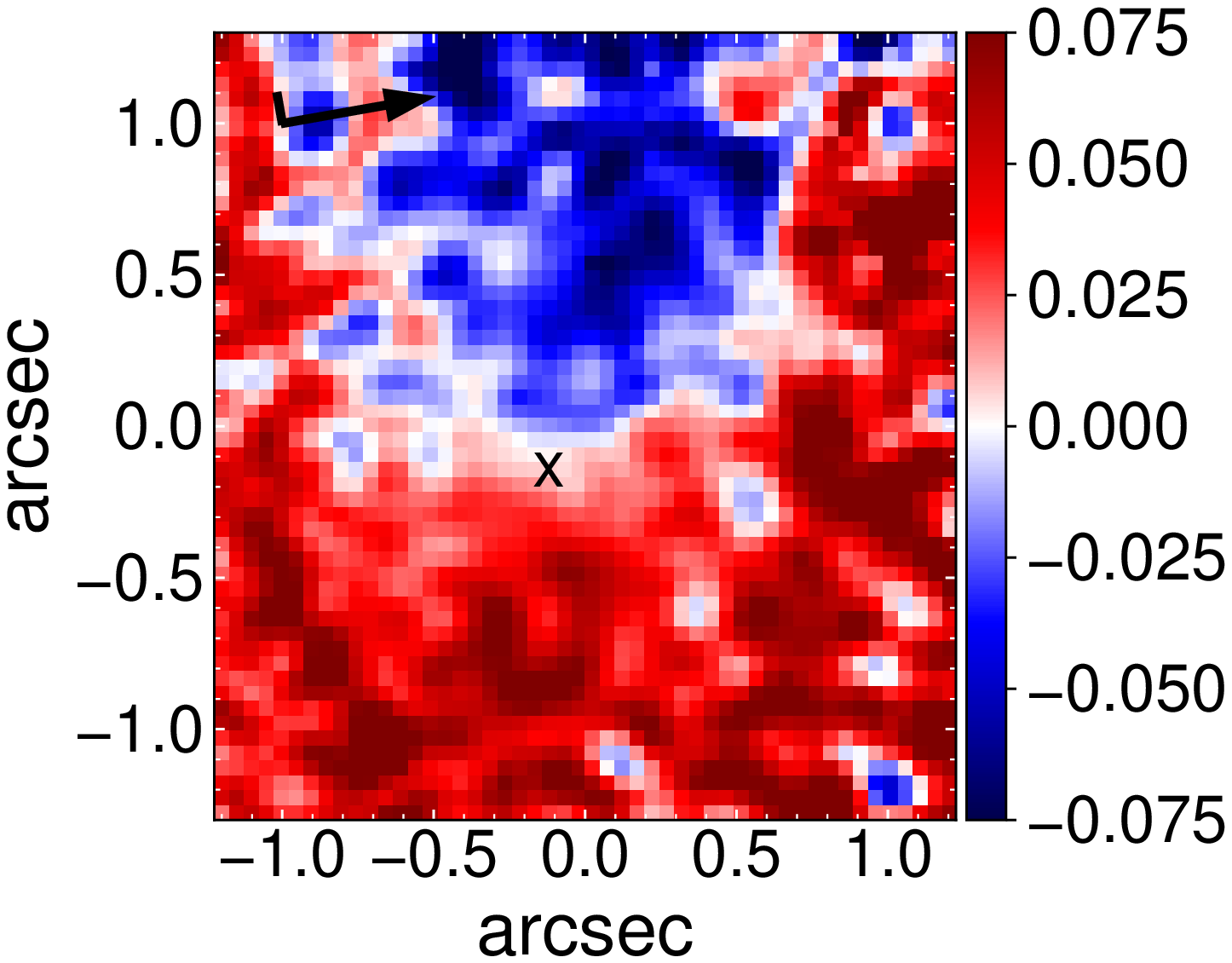}
    \includegraphics[scale=0.31]{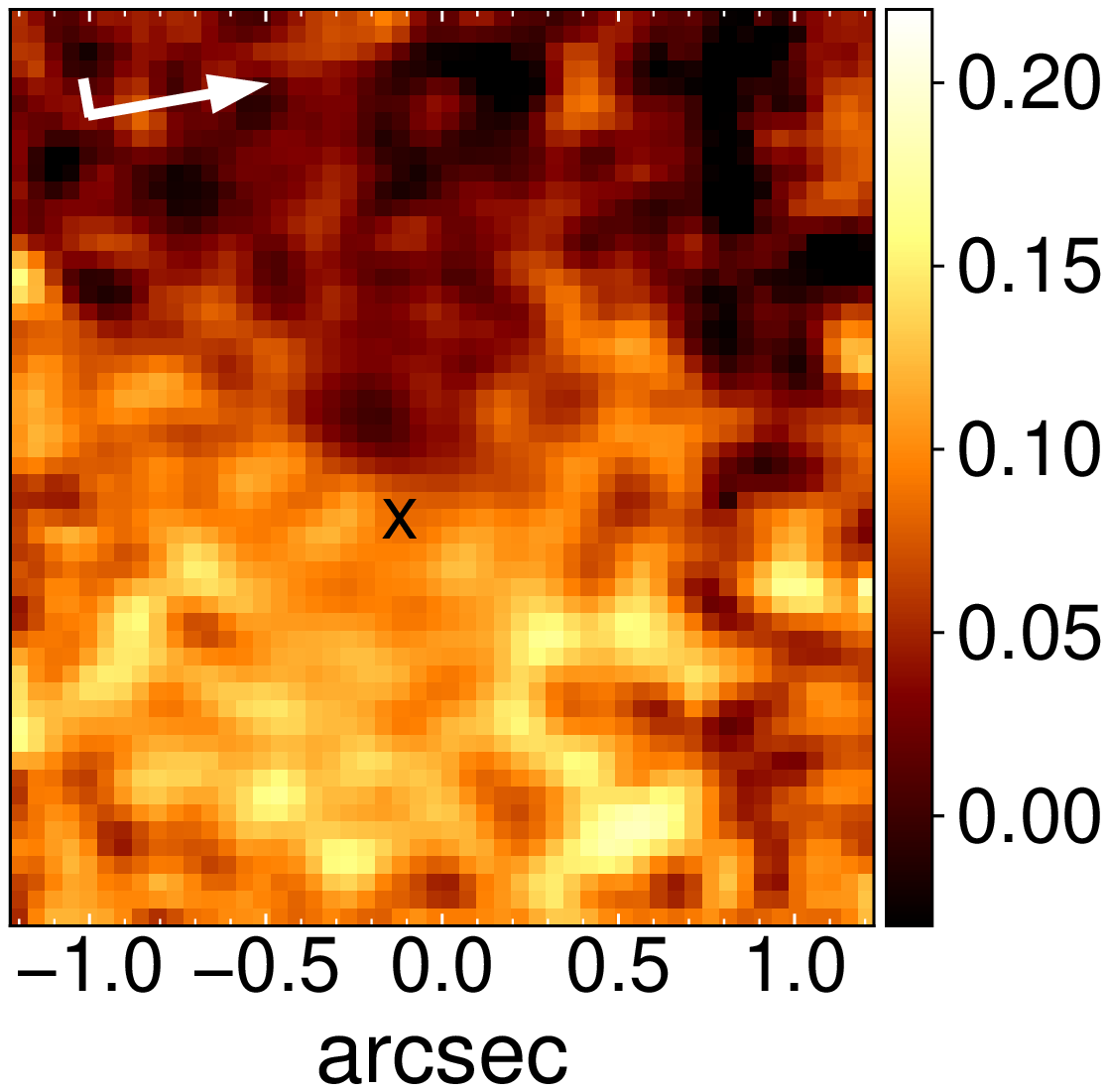}
    \caption{Stellar kinematic GH moments $h_3$ (left) and $h_4$ (right) of the central region of NGC 4546. \label{fig:kinematic_results_GH}}
\end{figure}

As a complement, we fitted again the LOSVD across the whole FOV of NGC 4546, but now using a GH function, i.e. including the third ($h_3$) and the fourth ($h_4$) moments in addition to the first two moments. We show only the results for $h_3$ and $h_4$ in Fig. \ref{fig:kinematic_results_GH}, since the $V_{RADIAL}$ and $\sigma_*$ maps extracted this way are very similar to the ones displayed in Fig. \ref{fig:first_kinematic_results}. One may see that the central region of NGC 4546 presents the typical anticorrelation between $V_{RADIAL}$ and $h_3$ that is usually seen in fast rotators (see e.g. \citealt{2011MNRAS.414.2923K}). The interpretation for the $h_4$ map, however, is not straightforward. It shows some degree of asymmetry, not typical in galactic nuclei. Comparing the $h_4$ and $\sigma_*$ maps, we noticed that the region located southwestward from the nucleus has higher $h_4$ values and slightly lower $\sigma_*$ results when compared to the north-east area of the galaxy. Such a result for the $\sigma_*$ parameter is seen in other measurements, as in long-slit data taken along the major axis of NGC 4546 \citep{1991MNRAS.248..544B} and also in the GMOS data cube of this object \citep{2016MNRAS.463.3860R}. These lower $\sigma_*$ values may be related to the presence of a kinematically cold younger stellar component \citep{2006MNRAS.369..853W}, whose evidences are revealed by means of stellar archaeology applied to the GMOS data cube of NGC 4546 \citep{ricci2013}. Another possibility may be related to the sigma-metallicity degeneracy. In this case, a template mismatch related to the metallicity of the stars in the stellar library and the metallicity of the galaxy may decrease the $\sigma_*$ measurements \citep{2008AN....329..968K}. It is possible that these higher values for $h_4$ are also related to one (or both) of these effects. However, a more robust test to check the cause of these results is beyond the scope of this paper.

\subsection{Central velocity dispersion} \label{sec:central_velocity_dispersion}

We calculated the central velocity dispersion $\sigma_{c}$ within 1 arcsec as
\begin{equation}
    \sigma_{c} \equiv \frac{\sum_{i=1}^{N_p}{I_i V_{RMS \\ i}}}{\sum_{i=1}^{N_p}{I_i}},
    \label{central_velocity_dispertion_equation}
\end{equation}
where $V_{RMS} = \left(V_{RADIAL}^2+\sigma_*^2\right)^{1/2}$, $I_i$ is the intensity of the stellar continuum in the spaxel $i$ and $N_p$ is the total number of spaxels within 1 arcsec. For this measurement, we used the observed $V_{RMS}$ map presented in Fig. \ref{fig:rms_maps}. The statistical error was calculated by means of a Monte Carlo simulation. We found $\sigma_c$ = 241$\pm$2 km s$^{-1}$ (3$\sigma$ confidence level). 

\section{The Jeans Anisotropic Models} \label{sec:jam}

To use the JAM equations that were proposed by \citet{2008MNRAS.390...71C}, we must assume that fast-rotating galaxies, as is the case of NGC 4546 \citep{2011MNRAS.414..888E}, have an axisymmetric distribution and that their velocity ellipsoids are aligned with the cylindrical coordinates $R$, $\phi$, and $z$ \citep{2007MNRAS.379..418C}, where $z$ is parallel to the axis of symmetry of the object. With those assumptions, the average second velocity moment in the line-of-sight $(\overline{v^2_{los}})$ of the stellar structure of a galaxy depends only on the stellar mass distribution, the black hole mass and on the anisotropic parameter $\beta_z$. In practice, the JAM technique calculates the stellar kinematics of an object by assuming only an axisymmetric shape and an inclination for the galaxy. 

\begin{figure*}
    \centering
    \includegraphics[scale=0.5]{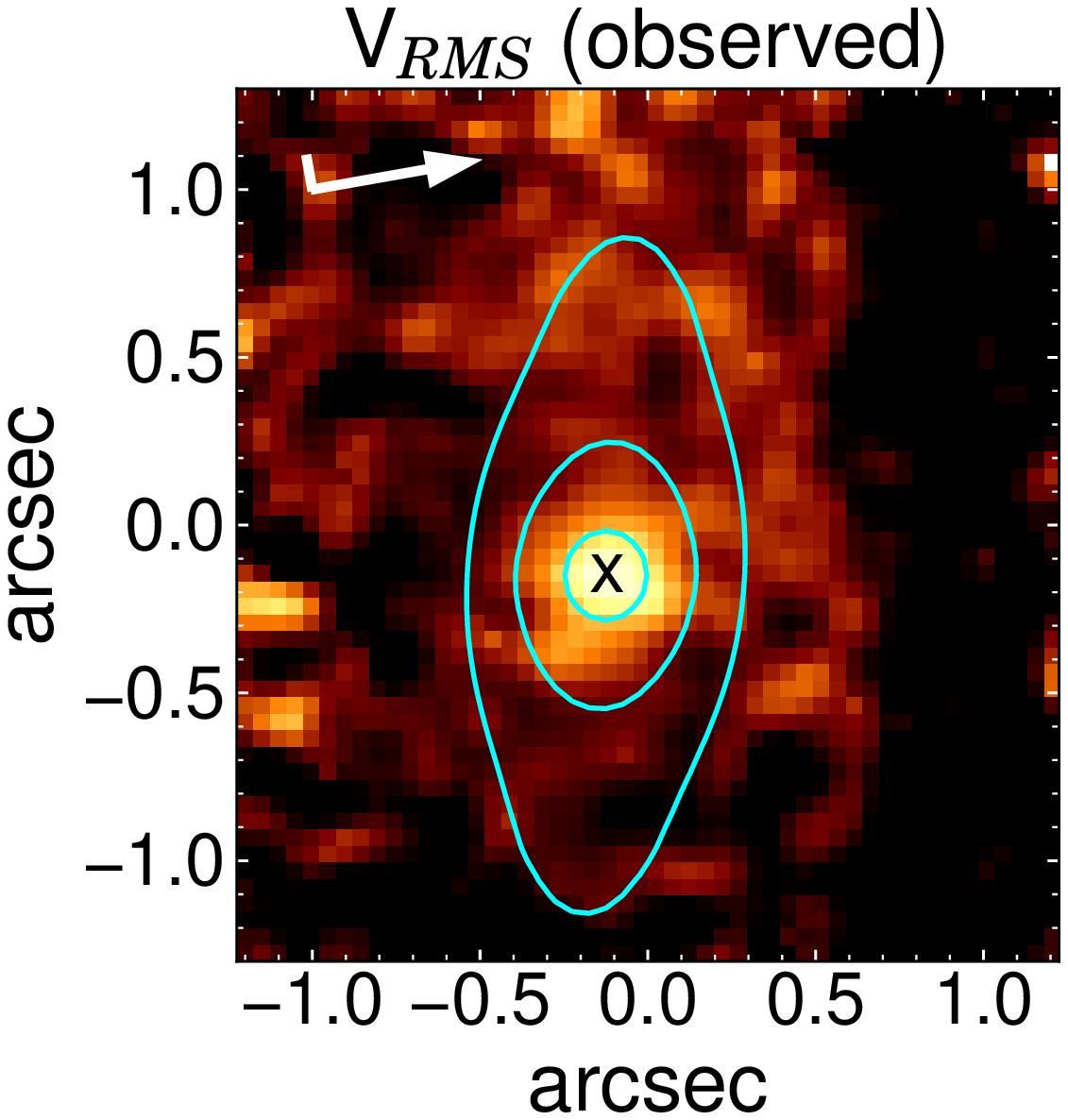}
    \includegraphics[scale=0.5]{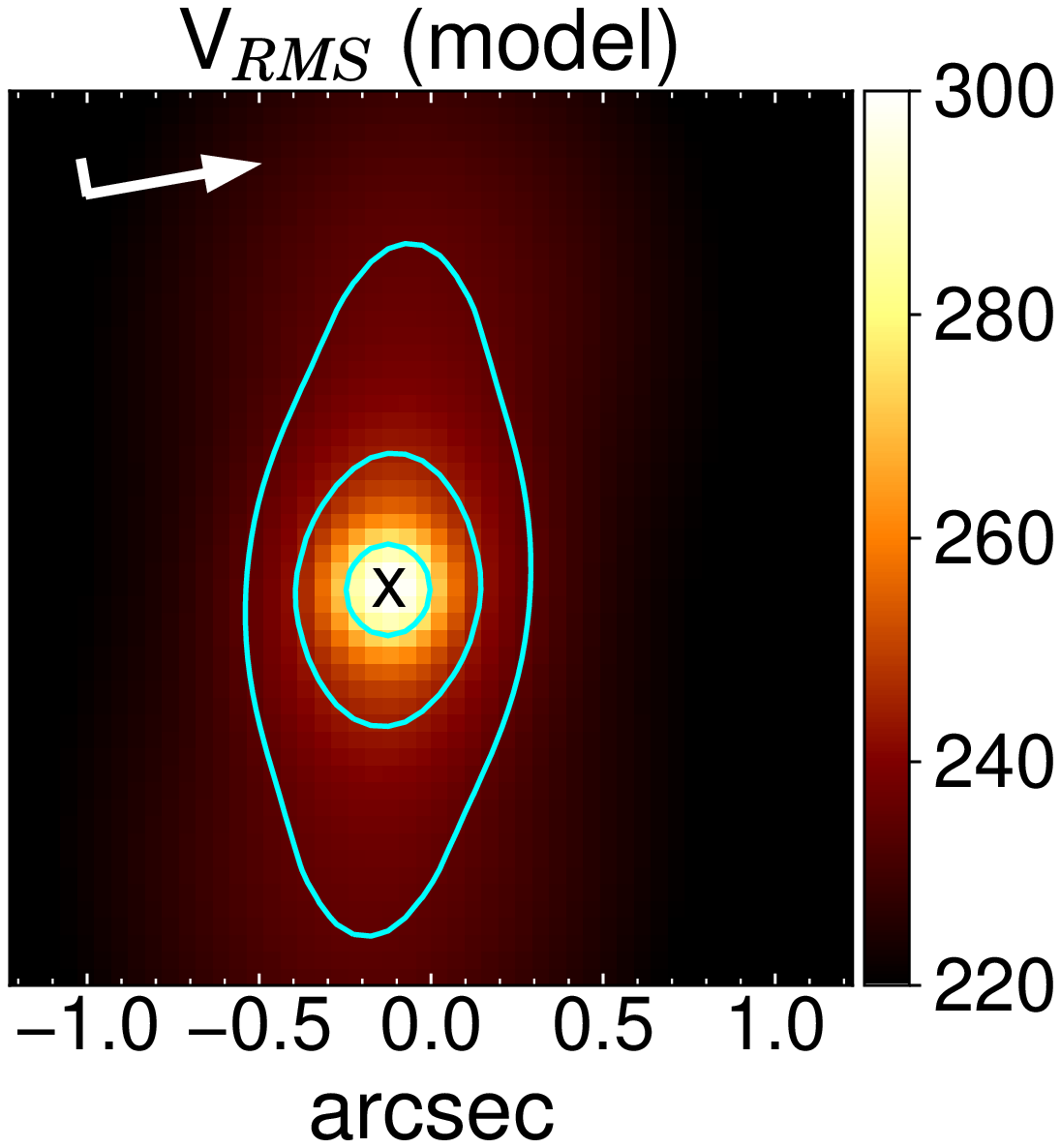}
    \includegraphics[scale=0.5]{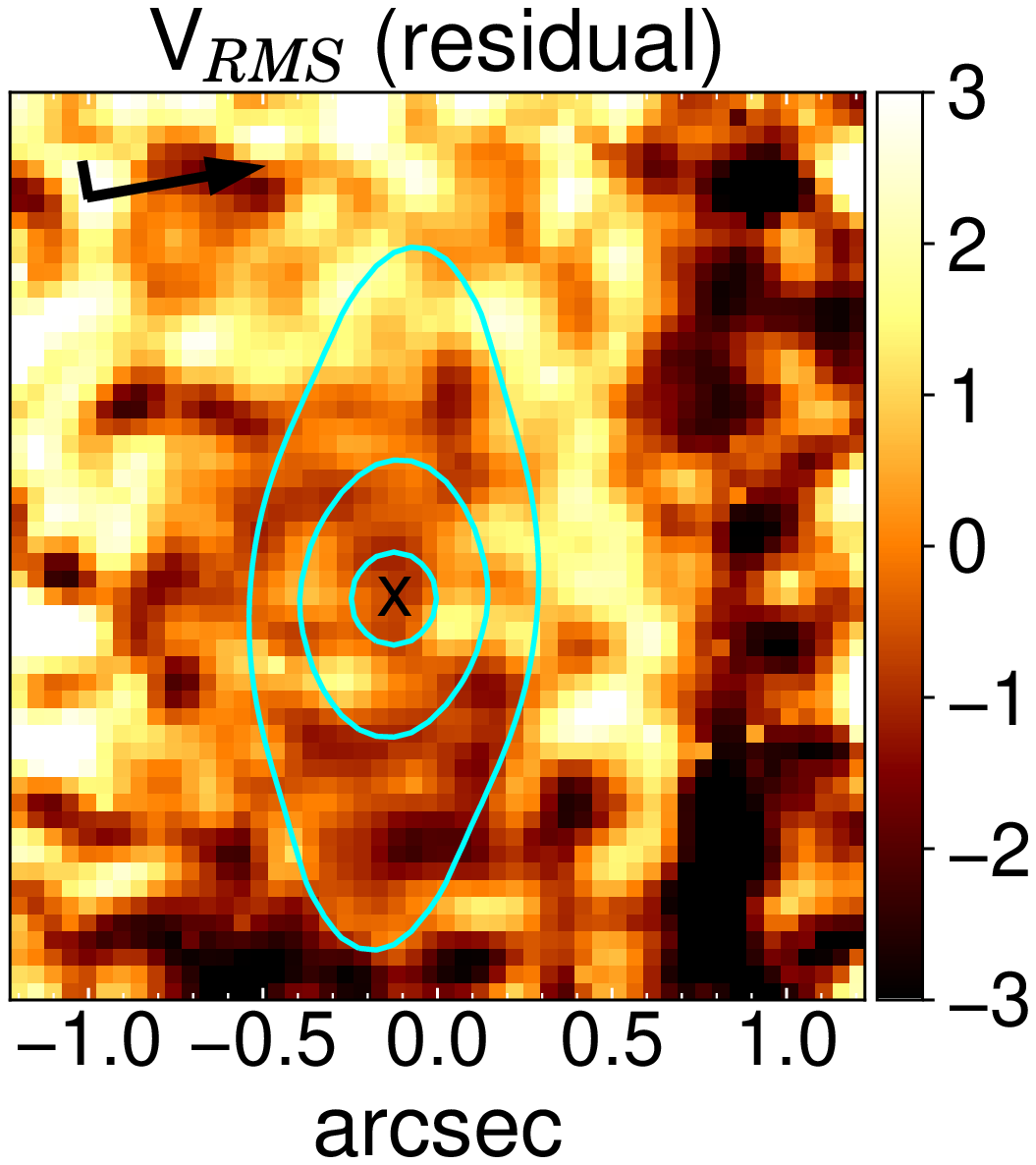}
    \caption{Left: observed $V_{RMS} = \sqrt{V_R^2+\sigma_*^2}$. In this case, both $V_{RADIAL}$ and $\sigma_*$ maps were obtained by fitting the absorption line profiles using a Gaussian function (i.e. setting the $h_3$ and $h_4$ parameters to zero in applying the {\sc ppxf} procedure). Centre: Best-fitting model for $V_{RMS}$ assuming $i$ = 69$^\circ$. Right: residual map divided by the error of the $V_{RMS}$ measurements. The first two maps are shown in km s$^{-1}$, while the map on the right ranges from --3$\sigma$ to 3$\sigma$. The contours shown in both images correspond to iso-velocity lines of 235, 245 and 280 km s$^{-1}$. \label{fig:rms_maps} }
    
\end{figure*}

The $\overline{v^2_{los}}$ parameter is well represented by the observed quantity $\left(V_{RMS}\right)^2$. According to \citet{2007MNRAS.379..418C}, this approximation is well suited when the kinematics is extracted from the data using only a Gaussian function to describe the profile of the LOSVD. The observed $V_{RMS}$ map is shown in Fig. \ref{fig:rms_maps}. The errors for $V_{RMS}$ were calculated through the propagation of the uncertainties found for $V_{RADIAL}$ and $\sigma_*$, resulting in $\Delta_{RMS}$ = 9 km s$^{-1}$.

The JAM implementation of \citet{2008MNRAS.390...71C} uses the results of the MGE expansion to calculate the mass distribution of a galaxy. The deprojection of an oblate axisymmetric galaxy using MGE corresponds to a sum of 3D Gaussian functions where the intrinsic axial ratio is associated with the inclination $i$ of the symmetry axis of the object with respect to the line of sight and also the observed axial ratio. Although the solutions of the deprojection of an axisymmetric galaxy are mathematically non-unique for all inclinations, except for an edge-on galaxy \citep{1987IAUS..127..397R, 1996MNRAS.279..993G}, the MGE results produce realistic intrinsic densities that are well suited for calculating the velocity moments of an object \citep{2008MNRAS.390...71C}. The conversion from luminosity density to a mass density is performed by calculating a mass-to-light ratio for each Gaussian function. In the case of NGC 4546, we make the simplistic assumption that the mass-to-light ratio is constant within the FOV of the data cube of this galaxy. 

\begin{figure}
    \includegraphics[scale=0.5]{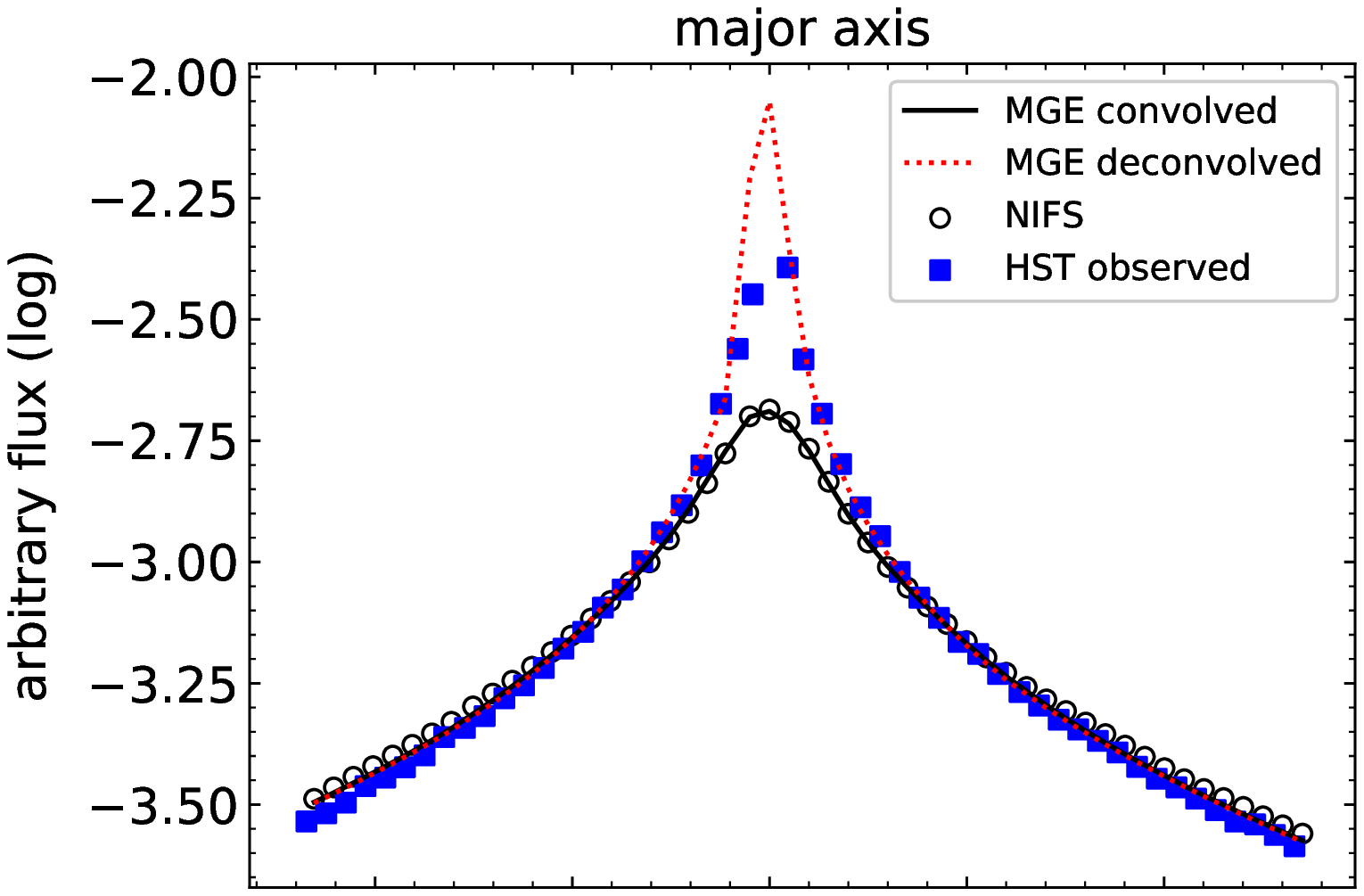}
    \includegraphics[scale=0.5]{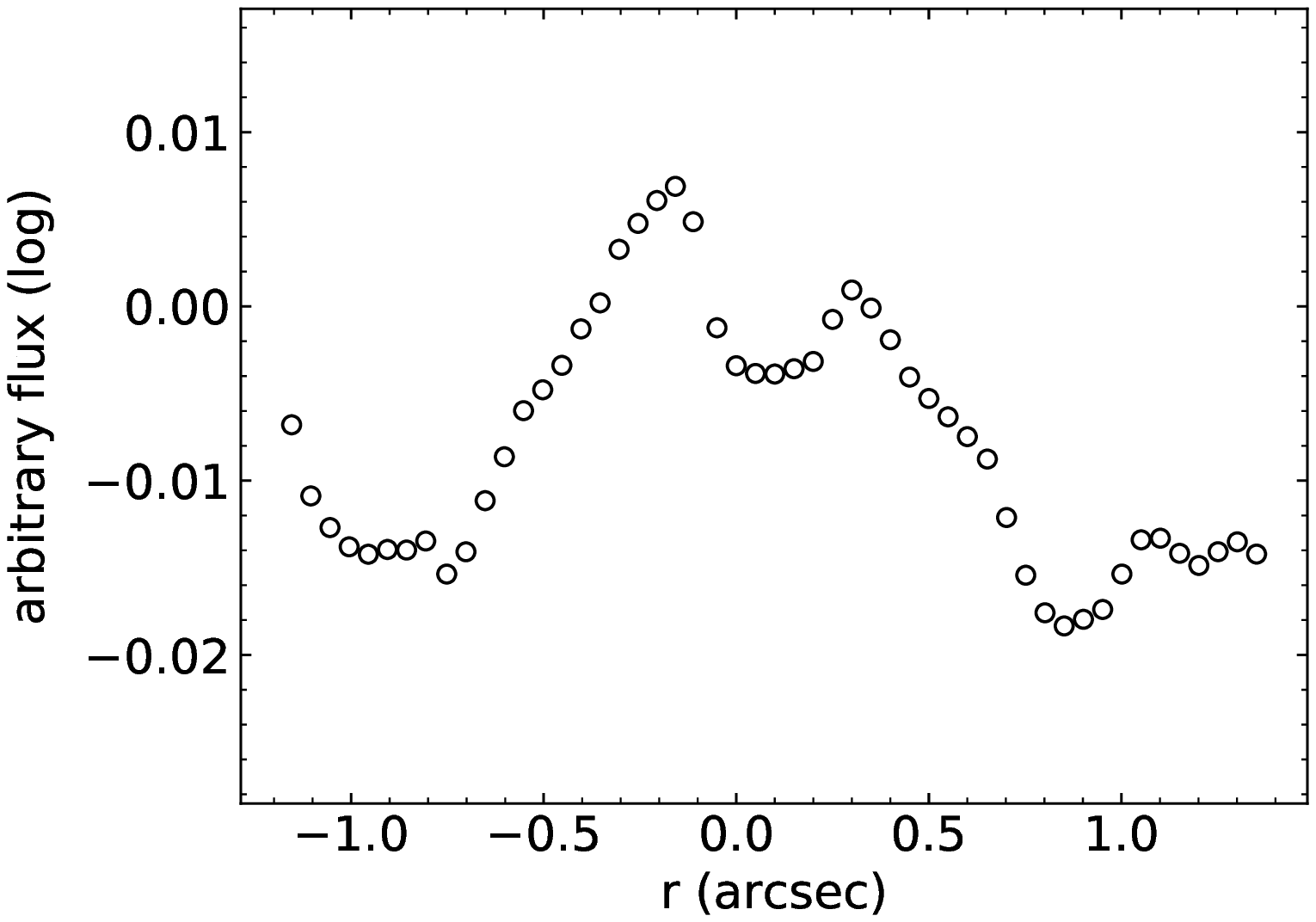}
    \caption{Top: surface brightness profiles along the major axis of NGC 4546. The black circles and the blue squares are related to the NIFS and the HST/WFPC2 observations, respectively (see Fig. \ref{fig:ngc4546}). The dotted red curve corresponds to the deconvolved MGE model and the full black curve describes to the MGE model convolved with the PSF of the NIFS data cube. Bottom: the ratio between the convolved MGE model and the observed NIFS data. One may see that the AGN of NGC 4546 does not affect the HST/WFPC2 profile of NGC 4546.  \label{fig:profiles_mge_hst_nifs}}
    
\end{figure}

In order to compare the JAM results with the observed data, it is necessary to convolve the models with the PSF of the NIFS data cube of NGC 4546. To determine this PSF, we followed the same procedure as in \citet{2009MNRAS.399.1839K}. In short, we assumed that the PSF of the NIFS data cube is well described by the sum of two concentric circular Gaussian functions, which are characterized by the FWHM of each element (FWHM$_N$ and FWHM$_B$ for the narrow and broad components, respectively) and by the intensity of the narrow component Int$_N$. The intensity of the broad Gaussian function Int$_B$ = 1 -- Int$_N$. We then convolved several PSFs with the deconvolved MGE model of the HST image of NGC 4546. We calculated several sets of PSFs, using 0.05 $\leq$ FWHM$_N$ $\leq$ 0.34 arcsec, 0.55 $\leq$ FWHM$_B$ $\leq$ 0.89 arcsec and 0.30 $\leq$ Int$_N$ $\leq$ 0.79 (total counts), with steps of 0.01 for all parameters in their respective units. The best-fitting PSF is the one that minimizes the $\chi^2$, obtained by comparing the result of the convolved MGE model with the average image extracted from the data cube between 2.24 and 2.26 $\mu$m, shown in Fig. \ref{fig:ngc4546}. We found FWHM$_N$ = 0.21 arcsec, FWHM$_B$ = 0.65 arcsec and Int$_N$ = 0.51 for the PSF of the NIFS data cube of NGC 4546. These results for the parameters of the PSF are shown in Table \ref{tab:journal_observations}. In Fig. \ref{fig:profiles_mge_hst_nifs}, we present the surface brightness profiles along the major axis of NGC 4546 extracted from the HST/WFPC2 and NIFS images shown in Fig. \ref{fig:ngc4546}, together with profiles of the deconvolved MGE model and of this same model convolved with the PSF of the NIFS data cube. One may see that the convolved MGE model fits the observed NIFS profile. With such a PSF, the predicted radius of the sphere-of-influence of the SMBH of NGC 4546 is resolved by the data (as shown in Section \ref{sec:introduction}, $R_{soi}$ = 18 pc, which corresponds to 0.26 arcsec at d = 14 Mpc). 

\begin{figure}
    \centering
    \includegraphics[scale=0.5]{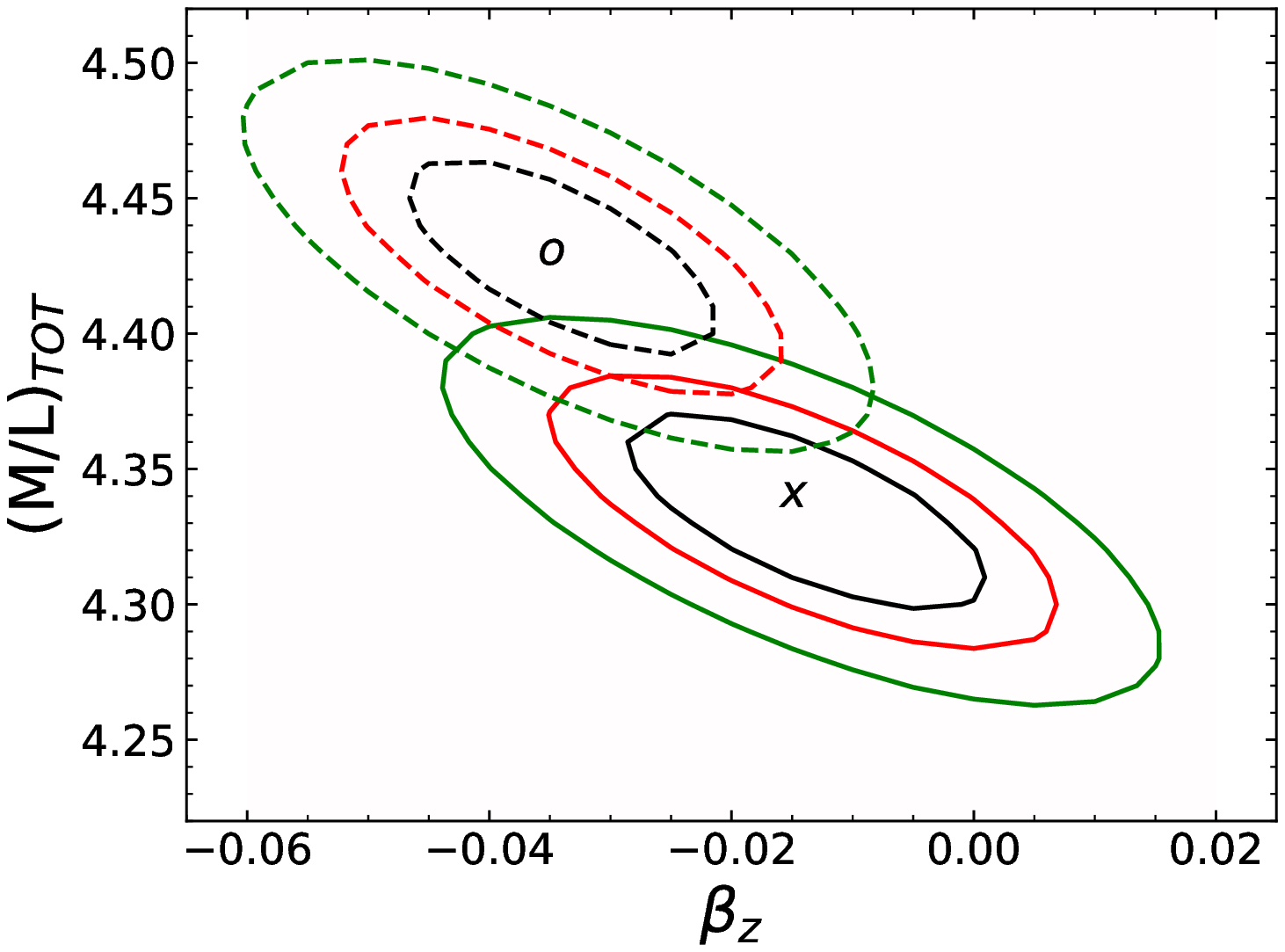}
    \includegraphics[scale=0.5]{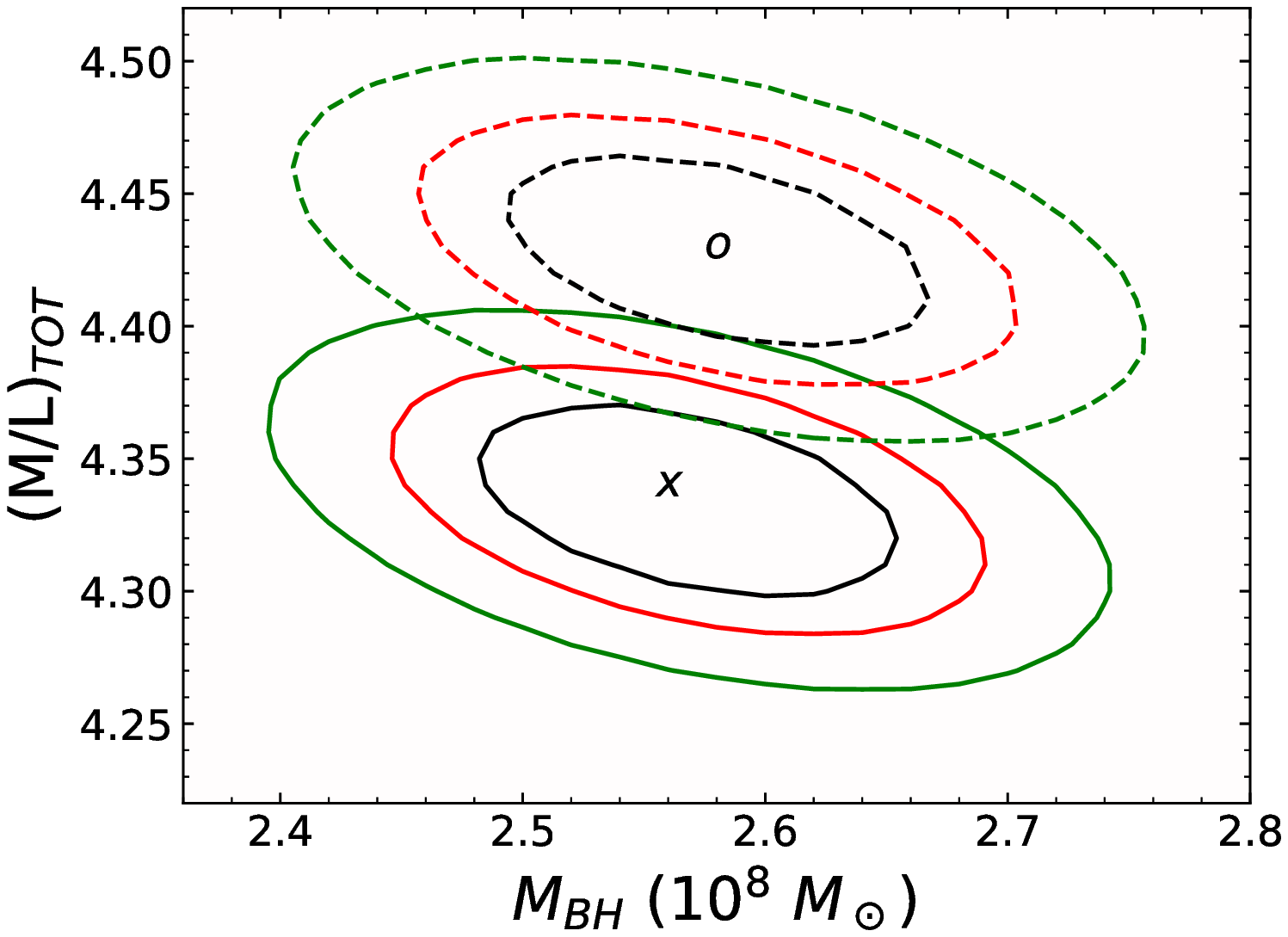}
    \includegraphics[scale=0.5]{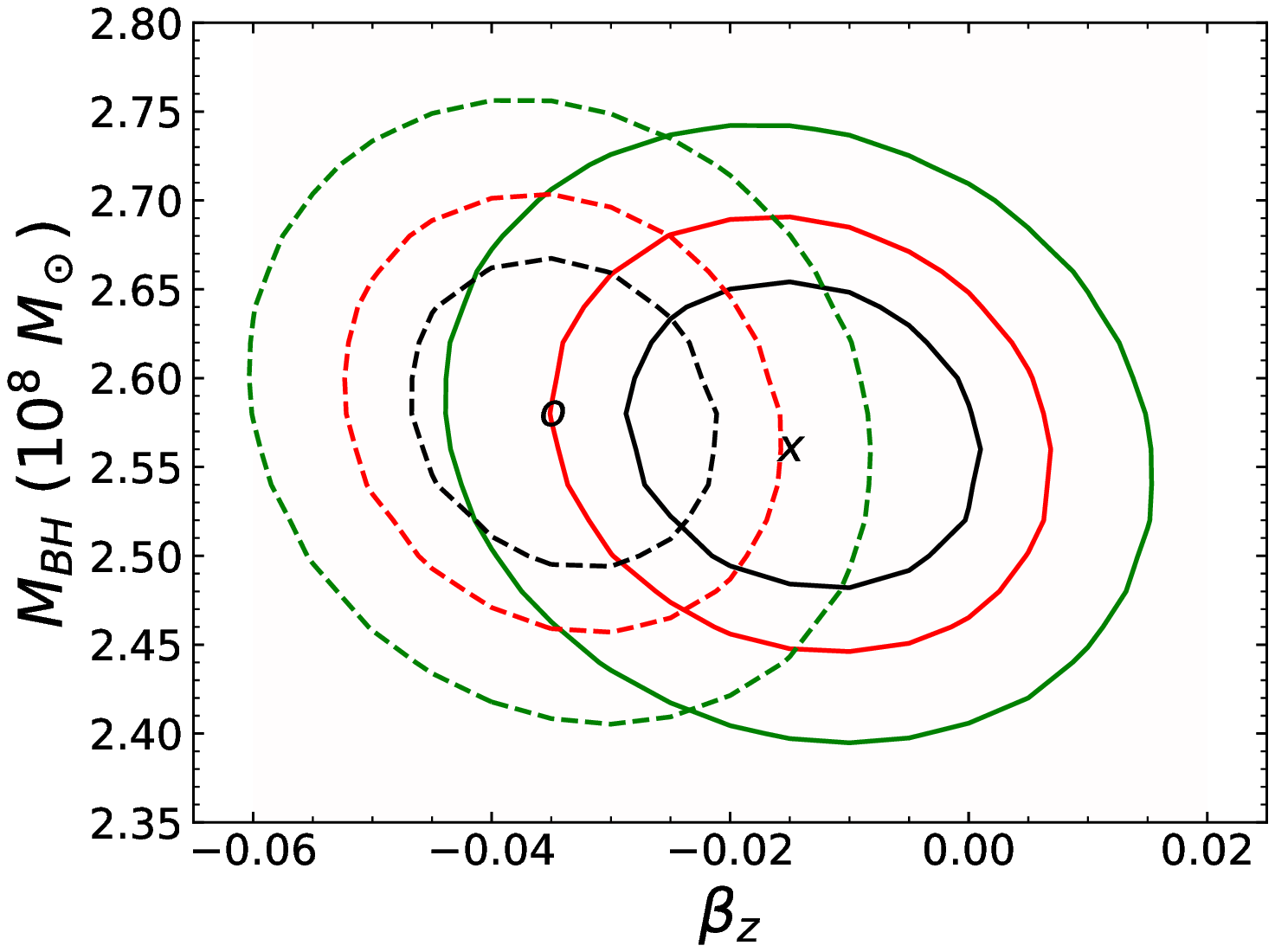}
    \caption{$\Delta \chi^2$ contours for different $(M/L)_{TOTAL}$, $ M_{BH}$, and $\beta_z$ values. The parameter spaces that were covered in the calculations were 4.24 $\leq (M/L)_{TOTAL} \leq$ 4.51, 2.36$\times$10$^8 \leq M_{BH}/M_\odot \leq$ 2.80$\times$10$^8$, and --0.065 $\leq \beta_z \leq$ 0.025, with respective steps of 0.01, 0.02$\times$10$^8$, and 0.005. Black contours correspond to 1$\sigma$ errors ($\Delta \chi^2$ = 2.3), red contours correspond to 2$\sigma$ errors ($\Delta \chi^2$ = 4.6) and green contours correspond to 3$\sigma$ errors ($\Delta \chi^2$ = 9.2). The full curves are related to the calculations assuming an inclination i = 69$^\circ$ and the dashed curves are associated with an edge-on model (i = 89$^\circ$). The $x$ and $o$ marks sets the position of the best-fitting models for i = 69$^\circ$ and for i = 89$^\circ$, respectively.}
    \label{fig:chi_contours}
\end{figure}

\begin{figure*}
    \centering
    \includegraphics[scale=0.5]{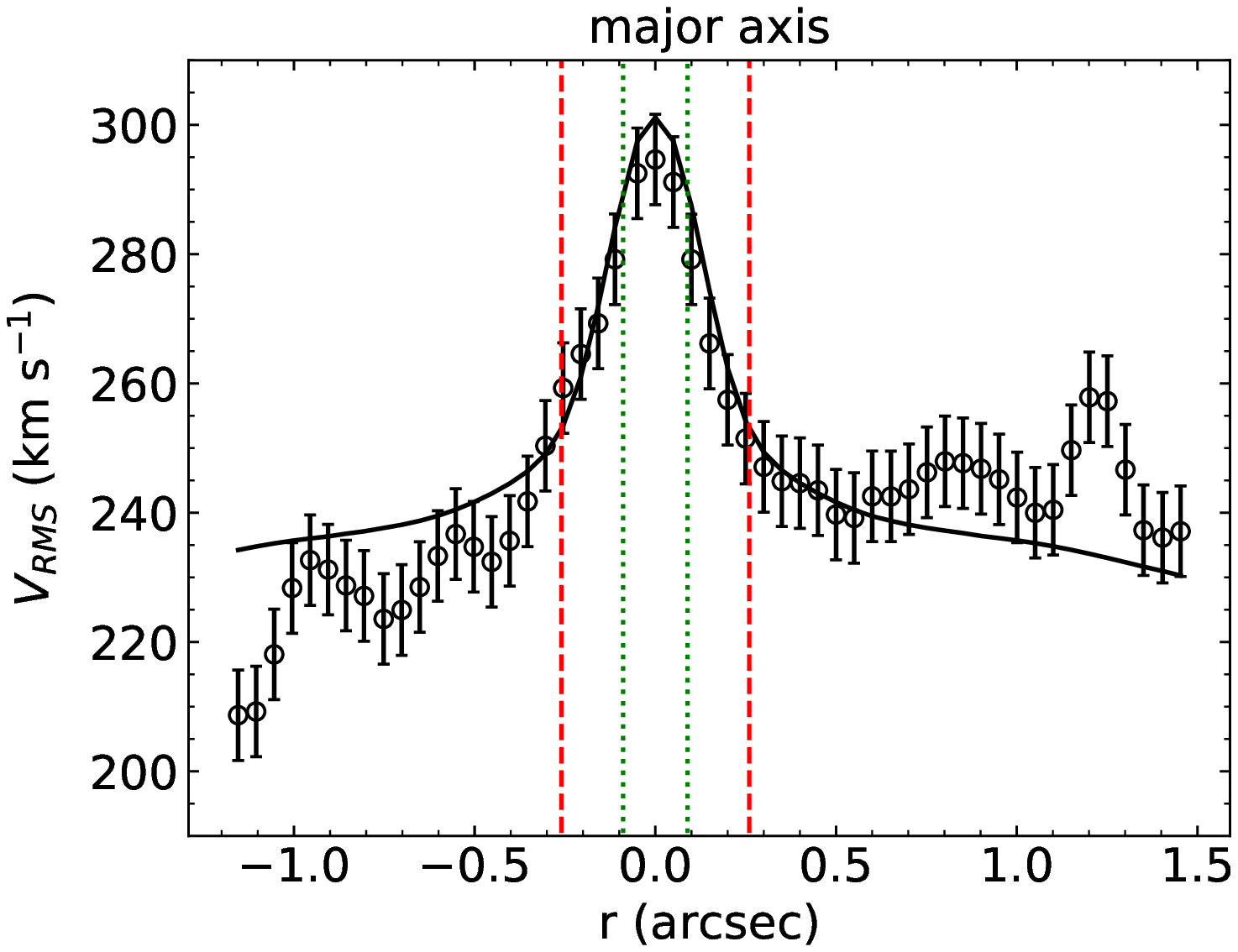}
    \includegraphics[scale=0.5]{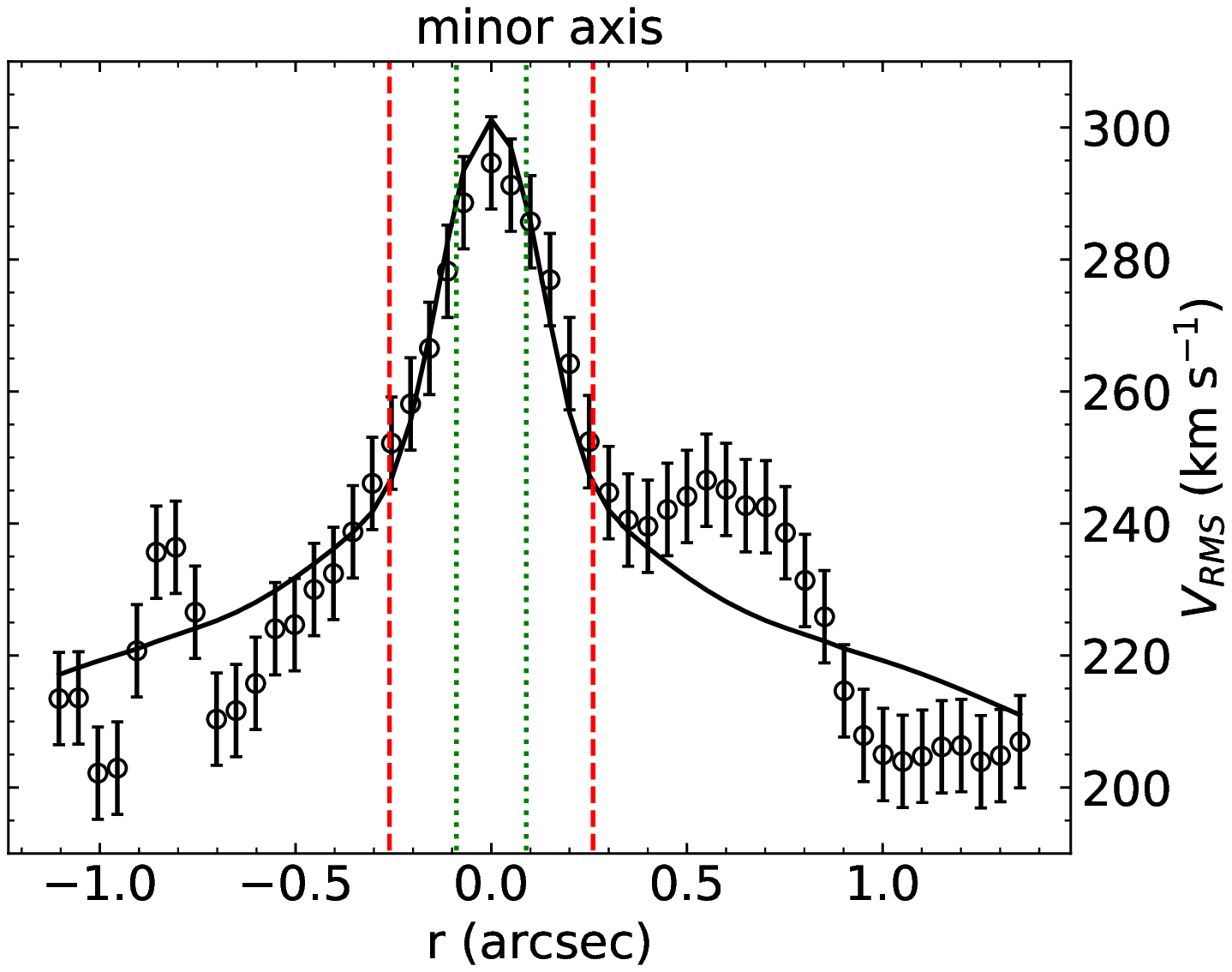}
   
    \caption{Profiles of the $V_{RMS}$ maps along the major (left, PA = 76$^\circ$) and minor (right) kinematic axis of NGC 4546. The hollow circles correspond to the observed values while the black curves are related to the best-fitting models of both parameters with $i$ = 69$^\circ$. The vertical dashed red lines correspond to the radius of the sphere of influece of the SMBH, while the vertical dotted green lines are related to the dispersion of the narrow component of the PSF ($\sigma_{PSF}$) of the data cube of NGC 4546 ($\sigma_{PSF}$ = FWHM$_N$(PSF)/2.3548).}
    \label{fig:profiles}
\end{figure*}

We built several maps for the $\left(\overline{v^2_{los}}\right)^{1/2}$ parameter assuming an inclination $i$ = 69$^\circ$, as found by \citet{2013MNRAS.432.1709C} using JAM applied to a SAURON data cube of NGC 4546. We also calculated $\left(\overline{v^2_{los}}\right)^{1/2}$ maps for an edge-on galaxy (in practice $i$ = 89$^\circ$) to minimise the influence of the mathematical degeneracy in the results of the deprojection procedure. We explored the parameter spaces using 4.24 $\leq (M/L)_{TOTAL} \leq$ 4.51, 2.36$\times$10$^8 \leq M_{BH}/M_\odot \leq$ 2.80$\times$10$^8$ and --0.065 $\leq \beta_z \leq$ 0.025, with respective steps of 0.01, 0.02$\times$10$^8$, and 0.005. In Fig. \ref{fig:chi_contours}, we present $\Delta \chi^2 = \chi^2 - \chi^2_{min}$ contours for different pairs of parameters. In these maps, $\Delta \chi^2$ = 2.3, 4.6, and 9.2 correspond to 1$\sigma$, 2$\sigma$ and 3$\sigma$ errors in a two-dimensional parameter space \citep{1976ApJ...210..642A}. We notice that the models with both inclinations are equivalent within the uncertainties. In particular, the black hole mass is not as affected by the inclination value as the other two parameters. The final uncertainty of a given parameter was obtained by marginalizing over the other three parameters. The best-fitting model for $i$ = 69$^\circ$ has $\chi^2_{min}$ per degree-of-freedom (DOF) = 1.010 and was obtained with $(M/L)_{TOT}$ = 4.34$\pm$0.07, $M_{BH}$ = (2.56$\pm$0.16)$\times$10$^8$ M$_\odot$ and $\beta_z$ = --0.015$\pm$0.030 (3$\sigma$ errors). The $\left(\overline{v^2_{los}}\right)^{1/2}$ map for this best-fitting model is shown in Fig. \ref{fig:rms_maps}. A profile comparing the observed $V_{RMS}$ and the calculated $\left(\overline{v^2_{los}}\right)^{1/2}$ along the major and minor kinematic axis of NGC 4546 is shown in Fig. \ref{fig:profiles}. 

\section{Possible systematics in the measurements} \label{sec:systematics}

In the previous sections, we presented only the statistical errors of all parameters. Here, we discuss a few issues that may cause systematic uncertainties in the determination of $(M/L)_{TOT}$, $M_{BH}$ and $\beta_z$. It is worth mentioning that the distance $d$ of the galaxy is a source of such uncertainty. Basically, $M_{BH}$ $\propto$ $d$ and $(M/L)_{TOT}$ $\propto$ $d^{-1}$. If one assumes $d$ = 19 pc \citep{2007A&A...465...71T} instead of $d$ = 14 Mpc \citep{2013AJ....146...86T}, then a factor of 1.36 applies to the best-fitting values of these two parameters.  

\subsection{Uncertainties related to the presence of an AGN} \label{sec:uncertainty_agn}

NGC 4546 has a type 1 LINER AGN in its centre \citep{2014MNRAS.440.2442R}, thus its light may affect the MGE fitting procedure that was performed in the HST image. However, the luminosity of the nuclear H$\alpha$ line is low (log L(H$\alpha$) = 39.3$\pm$0.3, \citealt{2015MNRAS.447.1504R}). In addition, there is no sign of a featureless continuum in the optical spectrum of this object  \citep{2014MNRAS.440.2419R}. Nevertheless, in order to quantify the contribution of the AGN with respect to the stellar component in the HST/WFPC2 image, we used the same GMOS data cube of this object that was presented in \citet{2014MNRAS.440.2442R}. We compared both the stellar and gas emission from the nuclear region of NGC 4546 after convolving the data cube with the transmission curve of the F606W filter from the HST/WFPC2 instrument. We found that the AGN contributes only with 2.4 \% of the total light in the nuclear region of NGC 4546 in this band. Moreover, the ratio between the MGE model convolved with the PSF of the NIFS data cube and the average image extracted from the NIFS data cube, shown in Fig. \ref{fig:ngc4546}, does not present any unresolved source in the nuclear region of NGC 4546 (see Fig. \ref{fig:profiles_mge_hst_nifs}). Thus, any error related to the emission of the AGN of NGC 4546 in the determination of the dynamical parameters of the stellar component of this galaxy is negligible. 

\subsection{Uncertainties related to the PSF} \label{sec:uncertainty_psf}

In this work, we assumed that the PSF of the NIFS data cube is given by a sum of two concentric Gaussian functions, as suggested by \citet{2009MNRAS.399.1839K}. We found that differences of $\sim$ 0.1 arcsec in both narrow and broad Gaussian functions and also of $\sim$ 10 \% in the intensity of the narrow component produce changes in the parameters that are consistent with the best values within the statistical errors. 

Another way to characterize the PSF of an NIFS data cube is by using the observations of the standard stars. This is not ideal, since there may be differences in the AO corrections between the science and the stars exposures or in the seeing of the observations \citep{2014MNRAS.438.2597M}. Nevertheless, to search for possible systematic errors associated with the PSF of the NIFS observations of NGC 4546, we decided to apply the MGE procedure on an image extracted from the data cube of one of the standard stars. A good fit was found with a set of four Gaussian functions. We then used this result of the PSF in the JAM procedure. Again, the discrepancy in all three parameters by using this solution for the PSF in the $\overline{v^2_{los}}$ models are within the statistical errors.

\subsection{Uncertainties related to the kinematic measurements} \label{sec:uncertainty_kin_measurements}

In order to check for systematic differences in our kinematic measurements, we compared our results with the ones obtained with the SAURON instrument for the ATLAS$^{3D}$ project \citep{2011MNRAS.413..813C}. By using their $V_{RADIAL}$ and $\sigma_*$ maps\footnote{Both the kinematic maps and the data cube from the ATLAS$^{3D}$ project were obtained in http://www-astro.physics.ox.ac.uk/atlas3d/} and also an image of the stellar continuum of their SAURON data cube, we calculated $\sigma_{c}$ within 1 arcsec using Equation \ref{central_velocity_dispertion_equation}. We found $\sigma_{c}$ = 235 km s$^{-1}$. This small discrepancy between the NIFS and SAURON measurements is probably caused by the differences between the spatial resolution of both data cubes. To test this, we calculated $\overline{v^2_{los}}$ maps using the best-fitting values of $(M/L)_{TOT}$, $M_{BH}$ and $\beta_z$ with characteristics of both instruments. For the SAURON data cube, we built this map with a spatial scale of 0.8 arcsec and we convolved it with a Gaussian function with a FWHM = 1.8 arcsec, which corresponds to the seeing of the observations of NGC 4546 made with this instrument \citep{2004MNRAS.352..721E}. Then, we used these model maps to calculate $\sigma_{cmodel}$. We found $\sigma_{cmodel}$ = 241 km s$^{-1}$ (NIFS characteristics) and 230 km s$^{-1}$ (SAURON characteristics). One may notice that a decrease of $\sigma_{c}$ is expected in the measurements performed in the SAURON data cube when compared to the NIFS data cube. Also, a qualitative comparison between the SAURON and NIFS maps of these kinematic parameters suggest that the measurements of both $V_{RADIAL}$ and $\sigma_*$ performed with the NIFS data cube are not affected by systematic errors.

\subsection{Uncertainties related to possible (M/L) variations} \label{sec:uncertainty_m_l_variations}

In this work, we assumed that $(M/L)_{TOT}$ is constant across the whole FOV of the NIFS data cube. Nevertheless, we investigate here if there are indications of radial variations of (M/L) that may affect, in particular, the measurement of the black hole mass of NGC 4546 (see e.g. \citealt{2019A&A...625A..62T}). Two possible reasons for this radial gradient are the presence of dark matter in the centre of NGC 4546 or variations in the stellar populations across this galaxy. \citet{2010MNRAS.408...97K} presented maps of the ages and metallicity of the stellar population of NGC 4546 using the SAURON data cube of this object. We noticed that in the FOV covered by the NIFS data cube, the variation of the stellar population with radius is negligible. However, we mentioned in Section \ref{sec:stellar_kinematics} that a young stellar component or a sigma-metallicity degeneracy may be related to the lower $\sigma_*$ values in a region located southwestward from the nucleus. This result is also seen in the profile of the $V_{RMS}$ map along the major axis (see Fig. \ref{fig:profiles}). Regardless of the reason for these lower $V_{RMS}$ values, we found that this issue may affect both $(M/L)_{TOT}$ and $M_{BH}$ measurements by $\sim$ 10 \%. 

JAM models applied to the same SAURON data cube using dark matter halos did not reveal any fraction of dark matter within one effective radius of this object \citep{2012Natur.484..485C, 2013MNRAS.432.1709C, 2013MNRAS.432.1862C}. Moreover, \citet{2013MNRAS.432.1709C} found $(M/L)_{TOT}$ = 4.34\footnote{As a matter of fact, \citet{2013MNRAS.432.1709C} presented their value of $(M/L)_{TOT}$ in the SDSS $r$ band. We applied a factor of 0.8 in order to convert this result to the Jonhson's R band, as suggested by \citet{2018MNRAS.477.3030K}. } within one effective radius of NGC 4546, which is the same value that we found within the NIFS FOV. This may be an indication that if a radial gradient of (M/L) caused by dark matter is present in NGC 4546, it is probably small and it will not affect the value we found for $M_{BH}$ of this object.

\subsection{Uncertainties related to the galaxy shape} \label{sec:uncertainty_galaxy_shape}

The deprojection of an axisymmetric galaxy using MGE depends on the Gaussian parameters ($I_j$, $\sigma_j$, and $q_j$) and also on the inclination $i$ of the $z$-axis with respect to the line of sight. For oblate objects, the flattening along the $z$ axis is associated with $i$ \citep{2002MNRAS.333..400C}. Moreover, any solution for the deprojection of a galaxy that is not edge-on is mathematically degenerate \citep{1987IAUS..127..397R, 1996MNRAS.279..993G}. As mentioned in Section \ref{sec:jam}, we assumed $i$ = 69$^\circ$ for NGC 4546 found by \citet{2013MNRAS.432.1709C} using JAM in the SAURON data cube. However, we also fitted the $V_{RMS}$ map with an edge-on model in order to minimise possible systematics associated with the degeneracy in the deprojection of the galaxy. One may note in Fig. \ref{fig:chi_contours} that the black hole mass is barely affected by the change in the inclination. The change in the $(M/L)_{TOT}$ and $\beta_z$ parameters are more pronounced, but still within the statistical errors (3$\sigma$). Thus, the degeneracy of the deprojection does not seem to affect the results obtained with JAM within the NIFS FOV. 

\citet{1991MNRAS.248..544B} reported the presence of a bar in NGC 4546, although \citet{2018ApJ...862..100G} and \citet{10.1093/mnras/staa392}, using recent observations of this object and more robust techniques on the photometric decomposition of galaxies did not find evidence for such a component. If present in NGC 4546, it does not invalidate our measurements, although it increases the degeneracy in the deprojection of the galaxy \citep{2012MNRAS.424.1495L} and it may also affect $(M/L)_{TOT}$ (and, consequently $M_{BH}$) by $\sim$ 15 \% \citep{2012MNRAS.424.1495L}, depending on the PA of the bar. 

\section{NGC 4546 on the $M_{BH}$ $\times$ $\sigma$ relation} \label{sec:m_sigma_relation}

\begin{figure}
    
    \includegraphics[scale=0.6]{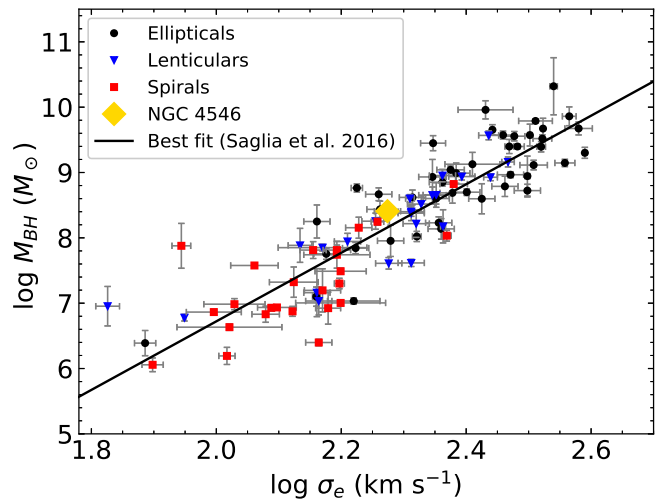}
    \caption{$M_{BH}$  $\times \, \sigma$ relation with the galaxy sample of \citet{2016ApJ...818...47S}, separated by elliptical, lenticular and spiral objects, together with the NGC 4546 results presented in this work. The best-fitting curve presented here was obtained by \citet{2016ApJ...818...47S} using all galaxies of the sample.}
    \label{fig:m_sigma}
\end{figure}

To verify whether our results for NGC 4546 follow the $M_{BH}$ $\times$ $\sigma$ relation, we used the velocity dispersion within one effective radius measured with the SAURON data cube of this galaxy ($\sigma_e$ = 188 km s$^{-1}$; \citealt{2013MNRAS.432.1709C}). In Fig. \ref{fig:m_sigma}, we compare our results with the values of $M_{BH}$ and $\sigma$ of the sample galaxies of \citet{2016ApJ...818...47S}, together with their best-fitting values for the $M_{BH}$ $\times$ $\sigma$ relation. We verify that NGC 4546, which is classified as an S0 object, is located in a region dominated by the lenticular galaxies of the \citet{2016ApJ...818...47S} sample. These results are in agreement with a formation scenario that included major wet mergers with gas accretion on the SMBH and also AGN feedback mechanisms \citep{2007ApJ...669...45H, 2013ARA&A..51..511K, 2016ApJ...818...47S}. Also, an isotropic velocity distribution in the central region of NGC 4546 is in accordance with what is seen in galaxies where major wet mergers were important \citep{2014ApJ...782...39T, 2018MNRAS.473.5237K}. It is worth mentioning that NGC 4546 has captured a nucleated dwarf galaxy within 2 Gyr ago \citep{2015MNRAS.451.3615N}, which is consistent with the fact that the stellar and gas components are counter rotating. 

\section{Conclusions} \label{sec:conclusions}

In this work, we presented new AO assisted NIFS observations from the central region of the lenticular galaxy NGC 4546. We measured the stellar kinematics (radial velocity and velocity dispersion) in order to perform a dynamical modelling of this component using the JAM technique, assuming that it has an axisymmetric shape. We fitted the stellar brightness distribution of NGC 4546 as a sum of 2D Gaussian functions using the MGE technique applied to an HST/WFPC2 image to calculate the mass distribution of this object. We summarize our main findings as follows: 

\begin{itemize}
    \item The central velocity dispersion of this galaxy (r $\leq$ 1 arcsec) $\sigma_c$ = 241$\pm$2 km s$^{-1}$ (3$\sigma$ confidence level). 
    \item The calculated total mass-to-light ratio in the R band $(M/L)_{TOT}$ = 4.34$\pm$0.07. 
    \item We found a black hole mass $M_{BH}$ = (2.56$\pm$0.16)$\times$10$^8$ M$_\odot$ (3$\sigma$ confidence level). This value is in agreement with what is expected for this galaxy using the $M_{BH}$ $\times$ $\sigma$ relation. 
    \item We found a value for the classic anisotropy parameter $\beta_z$ = --0.015$\pm$0.030 (3$\sigma$ confidence level), i.e. the central velocity distribution of NGC 4546 is isotropic. 
   
\end{itemize}

\section*{Acknowledgements}

Based on observations obtained at the Gemini Observatory, acquired through the Gemini Observatory Archive and processed using the Gemini IRAF package, which is operated by the Association of Universities for Research in Astronomy, Inc., under a cooperative agreement with the NSF on behalf of the Gemini partnership: the National Science Foundation (United States), National Research Council (Canada), CONICYT (Chile), Ministerio de Ciencia, Tecnolog\'{i}a e Innovaci\'{o}n Productiva (Argentina), Minist\'{e}rio da Ci\^{e}ncia, Tecnologia e Inova\c{c}\~{a}o (Brazil), and Korea Astronomy and Space Science Institute (Republic of Korea). This work is also based on observations made with the NASA/ESA Hubble Space Telescope, and obtained from the Hubble Legacy Archive, which is a collaboration between the Space Telescope Science Institute (STScI/NASA), the Space Telescope European Coordinating Facility (ST-ECF/ESA) and the Canadian Astronomy Data Centre (CADC/NRC/CSA). We have also made use of the NASA/IPAC Extragalactic Database (NED), which is operated by the Jet Propulsion Laboratory, California Institute of Technology, under contract with the National Aeronautics and Space Administration.

We also thank Davor Krajnovi\'c for a carefully reading of this manuscript, which improved a lot the quality of this work, and also Juliana Motter for English revision. TVR acknowledges CNPq for financial support under the grants 304321/2016-8 and 306790/2019-0. JES acknowledges FAPESP for financial support under the grant 2011/51680-6. Finally, we thank the anonymous referee for his/hers valuable comments. 




\bibliographystyle{mnras}
\bibliography{bibliography} 







\bsp    
\label{lastpage}
\end{document}